\documentclass{aa}  

\usepackage{graphicx}
\usepackage{txfonts}
\usepackage{lscape}
\usepackage[hidelinks]{hyperref}
\hypersetup{colorlinks=true, allcolors=blue
}
\begin{document}

   \title{Galaxy merger challenge: A comparison study between machine learning-based detection methods}

   \subtitle{}

    \author{B. Margalef-Bentabol\inst{1}\thanks{\email{B.Margalef.Bentabol@sron.nl}}\and 
          L. Wang\inst{1, 2}\and 
          A. La Marca\inst{1, 2}\and
          C. Blanco-Prieto\inst{3}\and
          D. Chudy\inst{4}\and
          H. Dom\'inguez-S\'anchez\inst{5}\and\newline
          A. D. Goulding\inst{6}\and
          A. Guzm\'an-Ortega\inst{7}\and
          M. Huertas-Company\inst{8}\and
          G. Martin\inst{9, 10}\and
          W.J. Pearson\inst{11}\and\newline
          V. Rodriguez-Gomez\inst{7}\and
          M. Walmsley\inst{12}\and
          R.W. Bickley\inst{13}\and
          C. Bottrell\inst{14}\and
          C. Conselice\inst{12}\and
          D. O’Ryan\inst{15}}

   \institute{SRON Netherlands Institute for Space Research, Landleven 12, 9747 AD Groningen, The Netherlands
   \and
   Kapteyn Astronomical Institute, University of Groningen, Postbus 800, 9700 AV Groningen, The Netherlands
   \and
    Centro de Astrobiolog\'ia (CAB), CSIC-INTA, Carretera de Ajalvir km4, 28850 Torrej\'on de Ardoz, Madrid, Spain.
   \and
   Astronomical Observatory of the Jagiellonian University, Faculty of Physics, Astronomy and Applied Computer Science, ul. Orla 171, 30-244 Cracow, Poland
   \and
   Centro de Estudios de Física del Cosmos de Aragón (CEFCA), Plaza San Juan, 1, E-44001 Teruel, Spain
   \and
   Department of Astrophysical Sciences, Princeton University, 4 Ivy Lane, Princeton, NJ 08544, USA
   \and
   Instituto de Radioastronom\'ia y Astrof\'isica, Universidad Nacional Aut\'onoma de M\'exico, Apdo. Postal 72-3, 58089 Morelia, Mexico
   \and
   Instituto de Astrofísica de Canarias, c/ Via Lactea sn, 38025 La Laguna, Spain
   \and
   Korea Astronomy and Space Science Institute, 776 Daedeokdae-ro, Yuseong-gu, Daejeon 34055, Korea,
   \and
   Steward Observatory, University of Arizona, 933 N. Cherry Ave, Tucson, AZ, USA,
   \and
   National Centre for Nuclear Research, Pasteura 7, 02-093 Warszawa, Poland
   \and
   Jodrell Bank Centre for Astrophysics, Department of Physics \& Astronomy, University of Manchester, Oxford Road, Manchester M13 9PL, UK
   \and
   Department of Physics and Astronomy, University of Victoria, Victoria, British Columbia V8P 1A1, Canada
   \and
   International Centre for Radio Astronomy Research, University of Western Australia, 35 Stirling Hwy, Crawley, WA 6009, Australia
   \and
   Department of Physics, Lancaster University, Bailrigg, Lancaster, LA1 4YB, UK}

  \date{Received -; accepted -}

% \abstract{}{}{}{}{} 
% 5 {} token are mandatory
 
  \abstract
  % context heading (optional)
  % {} leave it empty if necessary  
   {}
  % aims heading (mandatory)
   {Various galaxy merger detection methods have been applied to diverse datasets. However, it is difficult to understand how they compare. Our aim is to benchmark the relative performance of merger detection methods based on machine learning (ML).}
  % methods heading (mandatory)
   {We explore six leading ML methods using three main datasets. The first dataset consists of mock observations from the IllustrisTNG simulations, which acts as the training data and allows us to quantify the performance metrics of the detection methods. The second dataset consists of mock observations from the Horizon-AGN simulations, introduced to evaluate the performance of classifiers trained on different, but comparable data to those employed for training. The third dataset is composed of real observations from the Hyper Suprime-Cam Subaru Strategic Program (HSC-SSP) survey. We also compare mergers and non-mergers detected by the different methods with a subset of HSC-SSP visually identified galaxies.}
  % results heading (mandatory)
  {For the simplest binary classification task (i.e. mergers vs. non-mergers), all six methods perform reasonably well in the domain of the training data. At the lowest redshift explored $0.1<z<0.3$, precision and recall generally range between $\sim$70\% and 80\%, both of which decrease with increasing $z$ as expected (by $\sim$5\% for precision and $\sim$10\% for recall at the highest $z$ explored $0.76<z<1.0$). When transferred to a different domain, the precision of all classifiers is only slightly reduced, but the recall is significantly worse (by $\sim$20-40\% depending on the method). Zoobot offers the best overall performance in terms of precision and F1 score. When applied to real HSC observations, different methods agree well with visual labels of clear mergers, but can differ by more than an order of magnitude in predicting the overall fraction of major mergers. For the more challenging multi-class classification task to distinguish between pre-mergers, ongoing-mergers, and post-mergers, none of the methods in their current set-ups offer good performance, which could be partly due to the limitations in resolution and the depth of the data. In particular, ongoing-mergers and post-mergers are much more difficult to classify than pre-mergers. With the advent of better quality data (e.g. from JWST and Euclid), it is of great importance to improve our ability to detect mergers and distinguish between merger stages.}
  % conclusions heading (optional), leave it empty if necessary 
   {}

%When applied to real HSC observations, different methods agree well with visual labels of clear mergers but can differ by more than an order to magnitude in predicting the overall fraction of major mergers.

   \maketitle
%
%-------------------------------------------------------------------

\section{Introduction}\label{sec.introduction}

Galaxy mergers play a crucial role in galaxy formation and evolution in the hierarchical paradigm of structure formation \citep{White1978, Fakhouri2008, Conselice2014}. For example, mergers are expected to assemble stellar mass in addition to what is produced from star formation alone \citep{Rodriguez2016, Mundy2017, Duncan2019, Martin2021}, transform morphology \citep{Dubois2016, Rodriguez-Gomez2017, Martin2018, Martin2021},  kinematics \citep{Berg2014, Clauwens2018, Hani2018},  trigger starburst activity \citep{Cortijo-Ferrero2017, Pearson2019b, Patton2020}, and active galactic nucleus (AGN) activity \citep[][]{DiMatteo2012, Satyapal2014, Ellison2019}. However, the relative importance of mergers compared to other physical processes such as smooth gas accretion is still much debated \citep{Rodriguez2016, Fitts2018, McAlpine2020, Jackson2022}, and the details of how mergers are connected to specific phases along galaxy evolution histories (e.g. the triggering of the starburst and AGN phases) are not well understood \citep[][]{Martin2022}. One of the main challenges in better understanding the role of mergers in galaxy evolution is detecting them with sufficiently good reliability and completeness in a large enough sample across a wide redshift range. According to numerical simulations, a typical major merger between galaxies of similar masses can take several Gyr to complete \citep{Kitzbichler2008, Lotz2010, Husko2022}. A merger sequence includes both a pre-merger phase and post-merger phase to which several galaxy physical properties, such as star formation rate (SFR), are sensitive. The pre-merger phase is typically defined as the period during which the two gravitationally interacting galaxies approach each other, fly apart, and come close together again; the post-merger phase is defined as the period during which the colliding galaxies coalesce and form a single more massive galaxy \citep{Jiang2014, Snyder2017, Moreno2019}. The wide diversity in disturbed appearances and merging features (e.g. tidal tails, bridges, double nuclei) associated with different merger stages and merger types (e.g. with different mass ratios and gas content) along the relatively long merging sequence makes them difficult to characterise \citep{Martin2022, Desmons2023}. 

Traditionally, there are two categories of merger detection methods. The first relies on pair selection in which galaxies close on the sky and in redshift are identified as mergers \citep[e.g.][]{Woods2007, Ellison2013, Bustamante2020}. Ideally, spectroscopic redshifts are required to select genuine pairs. However, spectroscopic observations are very resource-intensive and time-consuming, and so photometric redshifts have been used in some cases \citep{Lopez-Sanjuan2015, Mundy2017, Duncan2019}, with increased contamination due to projection effect. Galaxy pairs can also suffer from flyby contamination (i.e. galaxies that interact briefly but do not merge in the end), even with the availability of spectroscopy. In addition, this method, by design, only selects pre-mergers when interacting galaxies are physically separated. The second set of methods involves identifying morphological disturbances in imaging, by visual inspection or using non-parametric morphological statistics, such as the CAS parameters (concentration, asymmetry, and smoothness; \citealp{Conselice2003}) or the combination of the Gini coefficient and $M_{20}$ (the second-order moment of the brightest 20\% of the light, \citealp{Lotz2004}). These methods require good-quality images in terms of spatial resolution and depth to identify merging features. However, even though CAS, Gini, and M-20 have been used to detect mergers \citep{Conselice2003, Lotz2004, Conselice2009, Mundy2017, Duncan2019},  they have been shown to yield significantly incomplete samples and can also contain disturbed galaxies that do not correspond to mergers \citep[e.g.][]{Pearson2019, Snyder2019, Bickley2021, Wilkinson2022}. One can also visually identify mergers, and the largest such dataset comes from Galaxy Zoo \citep{Lintott2008, Darg2010}. However, the main issues here are the difficulties in reproducibility and feasibility for large datasets.\footnote{Some attempts have been made to solve the reproducibility problem, but human classification is still needed \citep{Walmsley2022a}.} It can also be affected by low accuracy and incompleteness, particularly at high redshifts \citep{Huertas2015}. On the other hand, the visually most conspicuous mergers are expected to be very reliable.

The use of machine learning (ML) in astronomy has exploded in recent decades \citep[e.g.][]{Dieleman2015, Huertas-Company2018, Walmsley2020, Margalef2020, Zanisi2021, Karsten2023, Huertas2023}, and can be divided into traditional ML and deep learning (DL). Traditional ML algorithms have simpler structures (e.g. linear regression or decision trees) and typically rely on hand-crafted features to train  \citep{Domingos2012, Goulding2018, Martin2020, Lazar2023}. DL methods, based on artificial neural networks, are more sensitive to higher-order features. However, this comes at the cost of a much larger number of parameters, and they are thereby harder to train and require more data for optimisation \citep{pascanu2013, Schmidhuber2015, Goodfellow2016, tan2020}. Convolutional neural network \citep[CNN,][]{Fukushima1988, LeCun2015} architectures are a particular type of DL architecture that is extremely well suited for image classification. Their key features are convolutional layers that can identify patterns on different scales and extract relevant features, and a head of fully connected layers that perform the classification task \citep{oshea2015, Albawi2017}. In recent years, CNNs have shown great success in morphological galaxy classification, accurately reproducing visual labels \citep[][]{Dieleman2015, Huertas2015, Dominguez2018, Cheng2020, Walmsley2022a}. They solve problems such as reproducibility and applicability to large datasets. However, as they are designed to learn from visual labels in the training data, they can inherit biases from visual classification.

Deep learning has also been used for merger detection, with some studies successfully reproducing visual merger classifications \citep{Ackermann2018, Walmsley2019, Pearson2019, Pearson2022}. However, discerning mergers visually is much harder than discerning other more regular morphological classes (e.g. spirals and ellipticals), leading to incomplete and unreliable merger samples. To mitigate these issues, hydrodynamical simulations can be used to train DL algorithms. The main advantage of using simulations is the knowledge of ground truth on whether a galaxy is in the process of merging  (within a specific pre-defined time-frame). \cite{Pearson2019} trained a CNN on simulated galaxies from EAGLE \citep{Schaye2015}, processed to mimic the Sloan Digital Sky Survey (SDSS) observations. Their classifier achieved an accuracy of $65.2\%$ in the mock SDSS data and $64.6\%$ when applied to the real SDSS observations (compared to visual classifications). \cite{Bottrell2019} trained a CNN on mock SDSS galaxies generated from binary merger simulations \citep{Moreno2019} with the FIRE-2 physical model \citep{Hopkins2018}. The trained model was shown to achieve $87.1\%$ classification accuracy, discerning between isolated galaxies, pre-mergers, and post-mergers. \cite{Ciprijanovic2020} used a CNN to distinguish between mergers and non-mergers in simulated images from Illustris-1 \citep{Vogelsberger2014a, Vogelsberger2014b} at $z = 2$, reaching $79\%$ accuracy, which is slightly reduced to 76\% when noise mimicking real observations with the Hubble Space Telescope is added. \cite{Ferreira2020} trained a CNN on simulated galaxies at $z=0-3$ from IllustrisTNG \citep{Nelson2019},  which are further processed to resemble observations from the CANDELS survey. Using a Bayesian Optimisation,\footnote{Bayesian optimisation is a hyper-parameter optimisation method that takes into account past evaluations when choosing the hyper-parameter set to evaluate next, and therefore, focuses on areas of the parameter space that will most likely produce the best validation scores.} their model achieves $90\%$ accuracy when classifying mergers from the simulation, and can even distinguish between pre-mergers and post-mergers (with $87\%$ and $78\%$ accuracy, respectively). Finally, \cite{Bickley2021, Bickley2022, Bickley2023} trained a CNN for post-merger classification with images generated from IllustrisTNG and further processed to mimic the Canada France Imaging Survey \citep[][]{Ibata2017} for galaxies up to $z=1$, and achieved a classification accuracy of $88\%$. Even though many works on merger detection have focused on CNNs, traditional ML methods have also proven to be viable. \cite{Snyder2019} used non-parametric morphology statistics extracted from Illustris as features to train a random forest (RF) classifier. They achieved a completeness of $70\%$ at $0.5 < z < 3$, and purity ranging from $10\%$ at $z = 0.5$ to $60\%$ at $z = 3$. \cite{Guzman-Ortega2023} and \cite{Rose2023} obtained comparable performance at low and high $z$, respectively, with a RF classifier applied to IllustrisTNG. \cite{Nevin2019}, on the other hand, used a linear discriminant analysis in combination with non-parametric statistics to classify mergers in simulated images, achieving an accuracy of 85\% and precision of 97\% for major mergers.

All of these studies show that combining ML  with simulations is a promising approach for merger detection. However, these studies usually have very different set-ups. First, they may differ in the choice of simulation and galaxy formation physics. Thus, the impact of merging on galaxies may differ between simulations. Secondly, these studies differ in which observational survey they try to mimic (if any) and the chosen redshift range. Thirdly, they may differ in the definition of major mergers (with a mass or flux ratio of 1:4 or 1:3) or focus on different merger stages. Lastly, there are differences in the construction of major merger and non-merger samples, which generally should be complementary (i.e. galaxies are either in one category or another). However, this is often not the case as minor mergers with low mass ratios may be excluded from the non-merger sample. All of these differences make it very difficult to fully understand their relative performance. Another important aspect that has not been sufficiently studied is how well these methods trained on simulations perform on real observations;  there are few studies that focus on application to real surveys \citep[][]{Pearson2019b, Wang2020}. Some attempts were made to compare ML predictions with visual classifications, even though the latter does not necessarily represent the truth. Ideally, the source domain (data used to train the model) and the target domain (data to which the model will be applied) should be as similar as possible. The more different the two domains are, the less reliable the predictions will be in the target domain \citep{Bottrell2019, Ciprijanovi2020b}. One way to mitigate this is to generate simulated data that is as similar as possible to the observations. However, it is not always possible to fully recreate observational effects, and simulated galaxies may be intrinsically different from real galaxies. Therefore, models trained on simulations are expected to perform worse on observations \citep[][]{Dominguez2023}. 

In this paper we have several aims: (i) to apply leading ML-based methods to the same datasets, quantitatively comparing their performance on major merger identification; (ii) to assess whether the same performance is maintained when we apply classifiers trained on one simulation to another; (iii) to compare classifiers trained on simulations with visual labels for real observations. To achieve these goals, we utilised three different datasets. The first comes from the IllustrisTNG simulations, the second from the Horizon-AGN simulations, and the last from the Hyper Suprime-Cam Subaru Strategic Program (HSC-SSP)  survey \citep{Aihara2018}. 

The paper is structured as follows. In Sect. \ref{sec.data} we briefly describe the different datasets used in this work, including the two cosmological simulations of galaxy formation and evolution and real observations from the  HSC-SSP. In Sect. \ref{sec.challenge} we define the merger challenge and goals, explain how training and test datasets are created (including the whole process of generating mock images from simulations), and describe the metrics used for evaluation. In Sect. \ref{sec.methods} we outline the six merger identification methods explored in this work. In Sect. \ref{sec.results} we first compare the performance of the methods for the binary and multi-class merger classification tasks on the training data (TNG), and then we explore how the trained methods perform on a second set of simulations (Horizon-AGN). Finally, we investigate how they perform on real data and compare with visual labels. In Sect. \ref{sec.discussion} we present our conclusions and future directions.

\section{Data}\label{sec.data}

To study how different methods compare, we made use of two cosmological simulations {of galaxy formation and evolution (IllustrisTNG and Horizon AGN), and real observations from the HSC-SSP survey. The IllustrisTNG data were used to train all merger identification algorithms. The Horizon-AGN data were used for testing and quantifying how these methods perform when applied to a different dataset that was constructed in a similar way to the training data. This approach allows us to better understand how methods trained on simulations will behave when applied to observations. Below, we explain the main characteristics of the simulation and observation datasets used in this study.

\subsection{IllustrisTNG} \label{subsec.data.tng}

The IllustrisTNG project \citep{Nelson2019, Pillepich2018b, Springel2018, Nelson2018, Naiman2018, Marinacci2018},  is a series of cosmological magnetohydrodynamical simulations of galaxy formation and evolution that includes three runs spanning a range of volume and resolution, TNG50, TNG100, and TNG300, with comoving length sizes of 50, 100, and 300 Mpc $h^{-1}$, respectively. For this work, we used TNG300 due to the large number of galaxies it comprises and TNG100 to expand to lower-mass galaxies. The initial conditions for both runs are drawn from Plank results \citep{Plank2016}. Both runs follow dark matter (DM) particles,  gas cells, and stellar and supermassive black hole (SMBH) particles. TNG100 contains $1820^3$ DM particles with a mass resolution of $M_{\mathrm{DM,\ res}} = 7.5 \times10^6 M_{\odot}$, while TNG300 contains $2500^3$ DM particles with $M_{\mathrm{DM,\ res}} = 6 \times10^7 M_{\odot}$. The baryonic particle resolution is $M_{\mathrm{baryon\,res}}=1.4\times10^6$ M$_{\odot}$ and $M_{\mathrm{baryon\,res}}=1.1\times10^7$ M$_{\odot}$, for TNG100 and TNG300, respectively. Metal-enriched gas cools radiatively in the presence of a redshift-dependent, spatially uniform UV background. The cooling of gas is also affected by radiation from nearby SMBH. Gas above a density threshold of $0.1\ \mathrm{H}\ \mathrm{cm}^{-3}$ forms stars following the Kennicutt-Schmidt relation \citep{Kennicutt1998} and a Chabrier \citep{Chabrier2003} initial mass function (IMF). Stellar populations evolve via Type Ia supernovae, Type II supernovae, and asymptotic giant branch stars, returning mass and metals to the interstellar medium. Accreting SMBHs release energy via the `quasar mode' at high accretion rates with thermal feedback heating up gas surrounding the SMBH and via the `kinetic wind mode' at low accretion rates, producing SMBH-driven winds. We refer to \cite{Pillepich2018} for more details about IllustrisTNG. In this work, we only selected galaxies from the simulation snapshots 50-91, which correspond to redshifts from $z=1$ down to $0.1$. The time step between each snapshot is roughly $\sim$160\ \text{Myr} over this redshift interval. 

In the TNG simulations, DM halos are extracted using the friends-of-friends (FoF) approach \citep{Davis1985}, and only structures with more than 32 DM particles are considered DM halos. Substructures within the FoF groups (galaxies) are further extracted using a modified version of \texttt{SUBFIND}  algorithm \citep{Springel2001, Dolag2009}, which calculates the density field for all particles and cells, and only substructures with at least 20 resolution elements (stars and gas) are considered galaxies. Finally, the \texttt{SUBLINK} algorithm is used to construct the merger trees of galaxies, following the star particles and star-forming gas elements. Therefore, merger trees are constructed from baryon-based structures. This approach yields more accurate results than DM-based structures when studying mergers due to the baryon structures following the visible components, resulting in a closer observational definition. Furthermore, the merger time and mass ratio are calculated from the stellar masses of the two merging galaxies at the time when the secondary reached its maximum stellar mass \citep{Rodriguez2015}.

For TNG100, we selected galaxies with stellar mass $M_{*}>10^9M_{\odot}$ and for TNG300 with $M_{*}>8\times10^9M_{\odot}$, to ensure that most galaxies have a sufficient number of stellar particles (hence reasonably well resolved),  with the lowest mass galaxies in TNG100 ($M_{*}=10^9M_{\odot}$) consisting of $714$ particles, and in TNG300 ($M_{*}=8\times10^9M_{\odot}$, consisting of $727$ particles.

\subsection{Horizon-AGN}\label{subsec.data.horizon}

Horizon-AGN is a cosmological hydrodynamical simulation of galaxy formation and evolution \citep{Dubois14} with a comoving box size of 100 Mpc $h^{-1}$. The initial conditions are drawn from WMAP-7 cosmology \citep{Komatsu2011}. The total volume contains $1024^3$ DM particles with a mass resolution of $M_{\mathrm{DM,\ res}} = 8 \times10^7 M_{\odot}$ (similar to TNG300 but an order of magnitude lower than TNG100). The baryonic particle resolution is $M_{\mathrm{baryon\,res}}=2\times10^6$ M$_{\odot}$ (similar to TNG100 and better than TNG300). This simulation uses the adaptive mesh refinement code \texttt{RAMSES} \citep{Teyssier2002}, with a uniform grid that is refined down to a minimum cell size of 1 kpc constant in physical length. Gas cooling proceeds in the presence of a uniform UV background \citep{Haardt1996} via H, He and metal-enriched gas down to $10^{4}$~K \citep{Sutherland1993}. At densities above $0.1\ \mathrm{H}\ \mathrm{cm}^{-3}$, star formation proceeds with a fixed $2\%$ efficiency \citep{Kennicutt1998}. Chemical enrichment and kinetic energy injection into the gas is modelled via continuous stellar feedback from Type II SNe, Type Ia SNe, and stellar winds \citep{Leitherer1999, Girardi2000, Nomoto2007, Leitherer2010}. Additionally, SMBHs impart feedback on gas via the `quasar mode' (at Eddington ratios $\chi>0.01$), where thermal energy is injected isotropically into the surrounding gas with $1.5\%$ efficiency and the `radio mode' (at $\chi<0.01$), where kinetic energy is injected via bipolar outflows with jet velocities of $10^{4}\ \mathrm{km}\ \mathrm{s}^{-1}$. For more details on the physical processes in Horizon-AGN, we refer to \cite{Dubois14}. 

In the Horizon-AGN simulations, DM halos are identified using AdaptaHOP halo finder \citep{Aubert2004}. Only structures with more than 100 particles and with a density larger than 80 times the total matter density are considered DM halos. AdaptaHOP finder is also applied to the stellar distribution to identify galaxies with more than 50 particles. Merger trees of galaxies are built using the \texttt{TREEMAKER} algorithm \citep{Tweed2009}. Merger times and mass merger ratios are calculated when the minor companion is at its maximum mass (i.e. before it starts to lose any mass to the main galaxy), following the same approach described in \cite{Rodriguez2015}. Similarly to TNG, the merger trees are derived using baryon-based structures (in this case, stars) rather than DM ones.

We selected galaxies with $M_{*}>10^9$ (which ensures reasonably resolved galaxies, with a minimum of 500 particles for galaxies with $M_{*}=10^9$) and with redshifts between 0.1 and 1, to match galaxies selected from TNG.

Even though the two simulations use different methods for galaxy identification and to construct merger trees, we do not expect these differences to be significant, as they are both tracking the baryonic components of galaxies and use the same definitions of merger mass ratio. Moreover, \cite{Srisawat2013} compare different simulations and find that the choice of merger tree algorithms does not have a significant impact on the merger trees that they produce. In particular, \texttt{SUBLINK} and \texttt{TREEMAKER} (the codes used for TNG and Horizon-AGN, respectively) produce comparable merger trees. The choice of sub-grid physics applied to each observation could result in somewhat different galaxy populations (we explore the differences between the two simulations in Appendix \ref{sec.appensix.comparison.data}) since TNG is designed to reproduce some observed trends such as the galaxy mass size relation, galaxy stellar mass function and the SFR density \citep{Pillepich2018}, whereas in Horizon-AGN only the AGN feedback is implemented to reproduce the M-sigma relation \citep{Dubois14}. Despite these differences, both simulations find that the optical morphologies are in relatively good agreement with observations \citep[e.g. ][]{Rodriguez-Gomez2019, Dubois2016}.

\begin{figure*}
    \centering
    \includegraphics[width=1\textwidth]{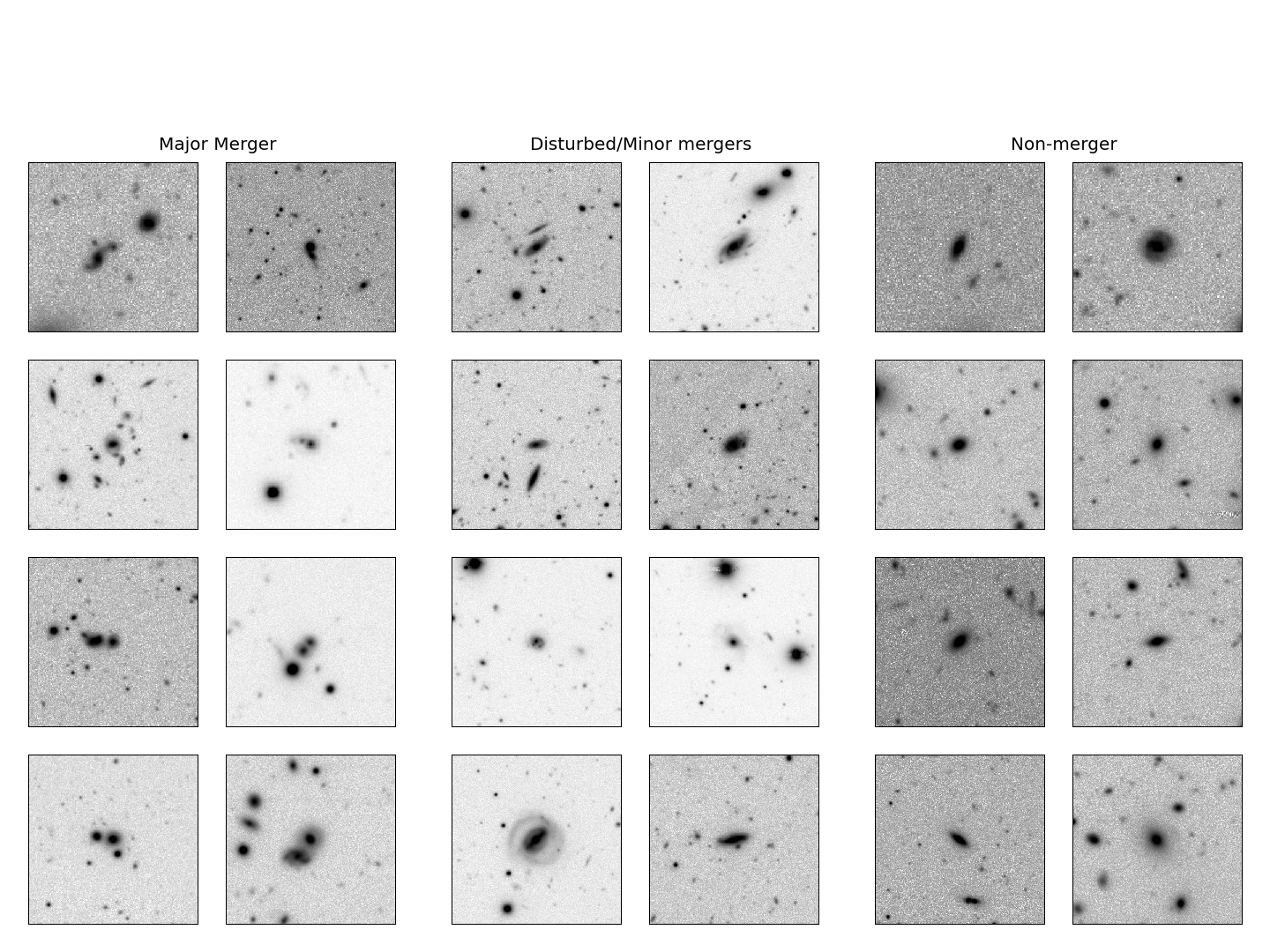}
    \caption{Example real HSC galaxies in the three visually classified groups: major mergers (two leftmost columns), disturbed or minor mergers (two middle columns), and non-mergers (two rightmost columns). The first two groups are from \protect\cite{Goulding2018}, and the last group is from this work. Images have a physical size of $\sim$160 kpc, displayed with an arcsinh inverted grey scale.}\label{fig.example_visual}
\end{figure*}

\begin{figure}
    \centering
    \includegraphics[width=0.5\textwidth]{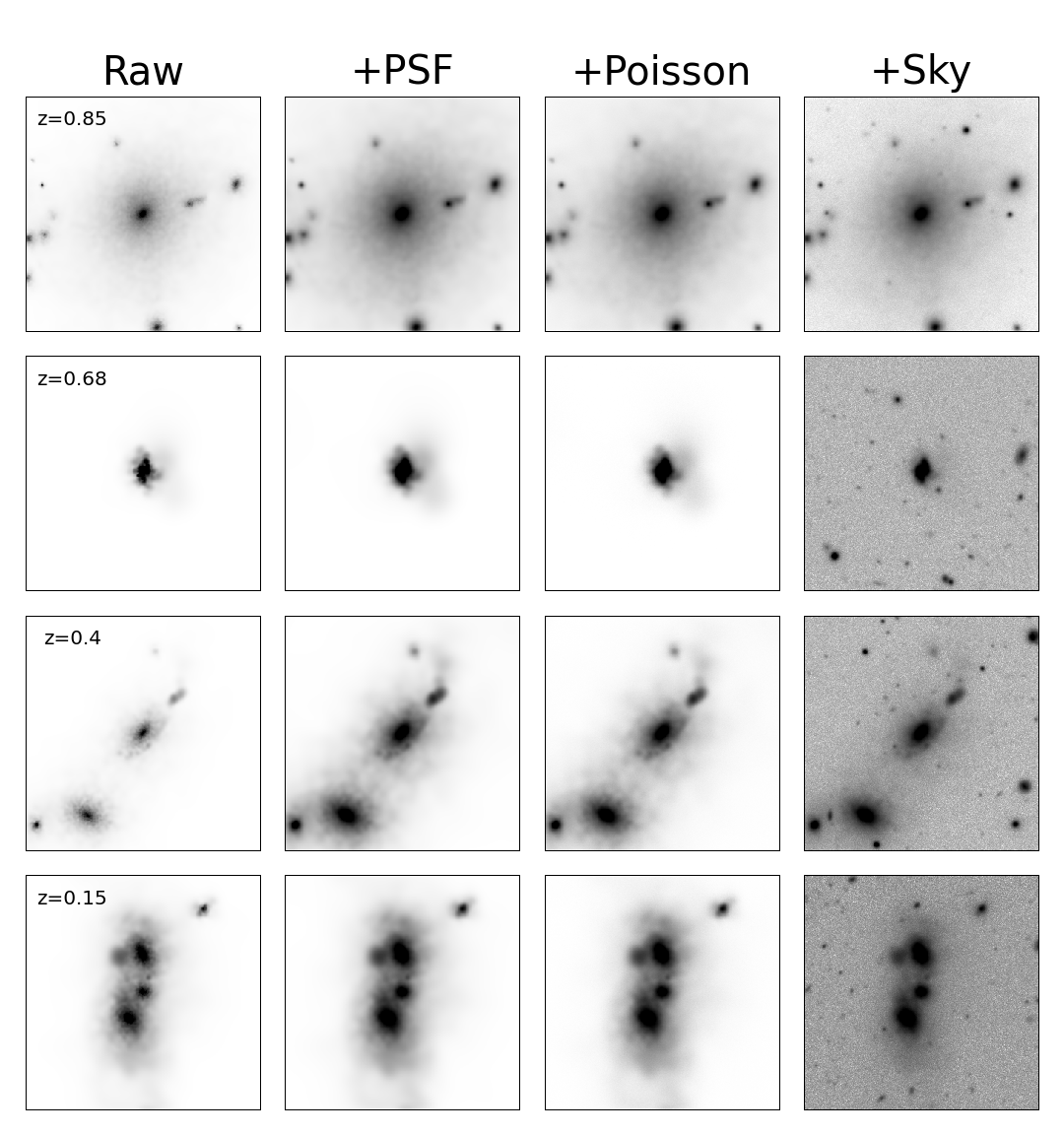}
    \caption{Steps used to create mock images for four randomly selected galaxies from TNG. From left to right are shown the raw simulated images, convolution with the HSC PSF, addition of Poisson noise, and injection into real sky background from HSC. Images have a physical size of 160 kpc, displayed with an arcsinh inverted grey scale.}\label{fig.mock_images_process}
\end{figure}

\begin{figure*}
    \centering
    \includegraphics[width=1\textwidth]{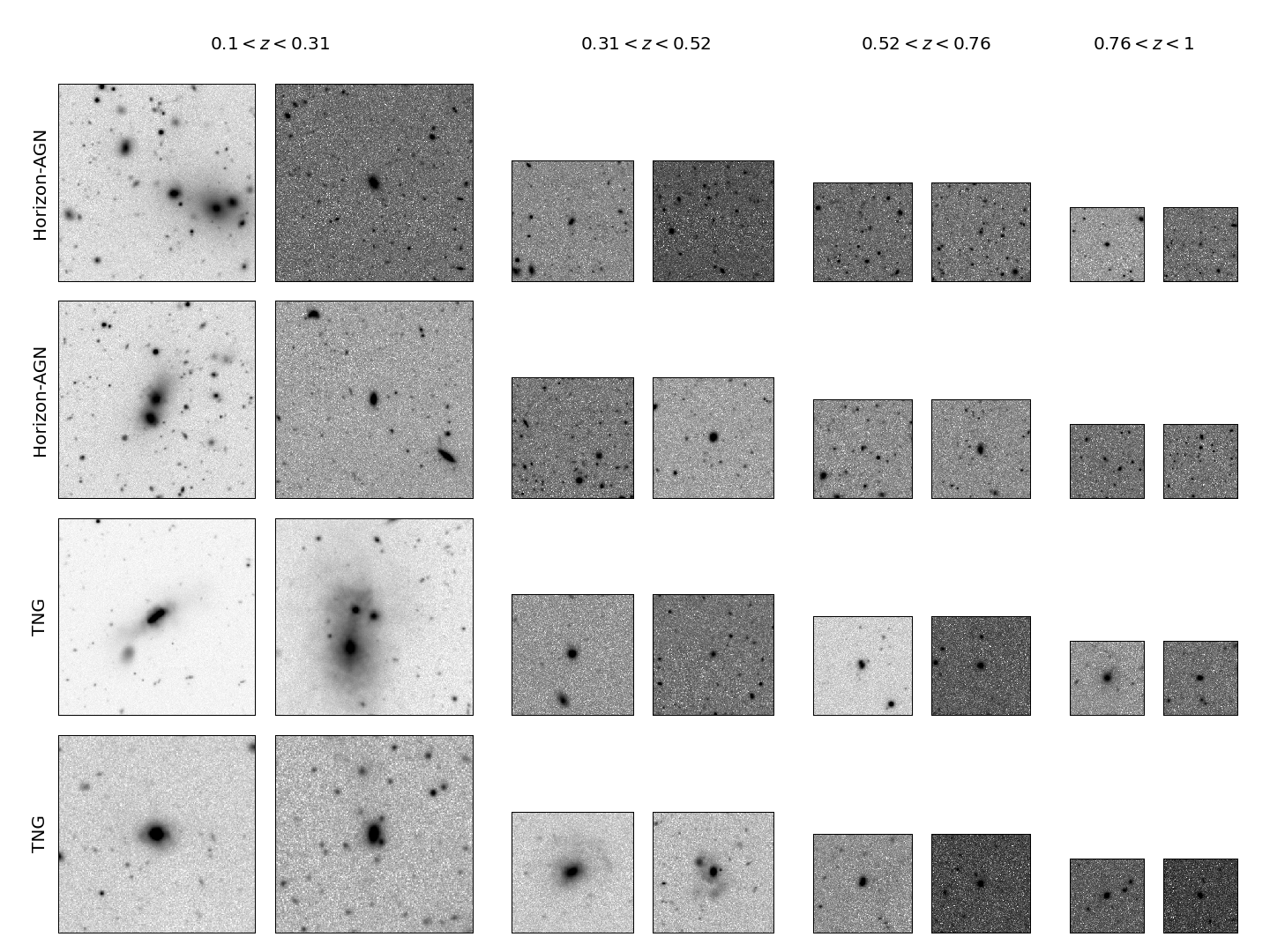}
    \caption{Example mock HSC images of simulated galaxies from TNG and Horizon-AGN. Images have sizes in pixels of 320 (at $0.1<z<0.31$), 192 (at $0.31<z<0.52$), 160 (at $0.52<z<0.76$), and 128 (at $0.76<z<1$), corresponding to $\sim$160 kpc at a given redshift, displayed using an arcsinh inverted grey scale.}\label{fig.mock_images_examples}
\end{figure*}

\subsection{HSC Subaru Strategic Program}\label{subsec.data.hsc}

The HSC-SSP survey is a wide-field optical imaging survey covering $\sim1200$ deg$^2$, conducted by the Hyper Suprime-Cam \citep[HSC; ][]{Miyazaki2018} imaging camera on the Subaru telescope, with a pixel resolution of 0.168 arcsec/pixel \citep[][]{Aihara2018}. We choose  the GAMA-09 field of $60\ \mathrm{deg}^2$ which spans $129^{\circ}\leq \mathrm{R.A.}\leq141^{\circ}$ and $-2^{\circ}\leq \mathrm{Dec.} \leq 3^{\circ}$ \citep{Liske2015}. We focused our study on the \textit{i}-band, given its depth of $\sim$26 mag at $5\sigma$ for point sources and seeing of $0.61''$ \citep[][]{Aihara2018}. Stellar masses and photometric redshifts are derived with the KiDS-VIKING \citep{Kuijken2019, Edge2013} photometry (see \citealp{Wright2019} for details). In summary, photometric redshifts (or spectroscopic redshifts when available) are derived from the KiDS-VIKING 9-band photometry using the Bayesian Photometric Redshift code \citep[BPZ,][]{Benitez2000}. The normalised median-absolute-deviation achieved is $(z_{phot} - z_{spec})/(1 + z_{spec})=0.061$. We then constructed our HSC sample by randomly selecting $\sim$120000 galaxies in GAMA-09, matching the redshift ($0.1<z<1$) and stellar mass range ($M_{*}>10^9M_{\odot}$) of the simulated samples. Stellar masses are estimated using the template fitting code \textsc{le phare} \citep{Ilbert2006}, which models the photometry with a library of stellar population models of \cite{Bruzual2003}, \cite{Chabrier2003} IMF, \cite{Calzetti1994} dust extinction law and exponentially declining star formation histories. We downloaded the $i$-band coadd images of our sample from the HSC-SSP Data Release 3 \citep{Aihara2022} using the DAS cutout facility. 

We included a further $\sim$2000 galaxies with visual classification labels from \citep{Goulding2018}. These galaxies come from an initial random sample of 5900 star-forming galaxies (according to their position in the UVJ diagram) and have been visually classified using K-corrected 3-colour images of size $50\times50\ \mathrm{kpc}$. These $\sim$2000 galaxies with visual classification labels fall into one of the two classes:

\begin{itemize}
    \item Visual major mergers: Strongly interacting massive galaxy pairs or post-mergers, including galaxies that have double nuclei. In the case of clear evidence of interaction with a distinct companion galaxy, only systems in which the flux ratio between the two interacting galaxies is  $>$ 1:4 are included.
    \item Disturbed--Minor-mergers: Galaxies that do not have clear signs of a major merger but show irregular, disturbed, asymmetrical, or torqued morphologies, along with galaxies considered to be minor mergers (galaxies with a companion but with a flux ratio between the systems $<$ 1:4).
\end{itemize}

Additionally, we created a visually identified sample of clear non-mergers, which are galaxies with no signs of merging or any type of disturbed morphologies. From our HSC sample, we randomly classified galaxies and selected those that are deemed as clear non-mergers to create this sample of $\sim$1000 galaxies to match the number of galaxies in the previous two groups (visual major mergers and disturbed--minor mergers). Examples of the three visually classified groups are shown in Fig. \ref{fig.example_visual}.

\section{The merger challenge}\label{sec.challenge}

\subsection{Goals}\label{subsec.challenge.goasl}

One of the main purposes of building a merger classifier is to apply it to real observations to measure merger fractions and merger rates. However, it is impossible to quantify the performance that a method trained on simulations has on real observations, at least not in the same way we assess such performance in the simulations. This is because we do not have access to the ground-truth merger history of real galaxies. Instead, the classification labels that we rely on in observations can be subjective, highly uncertain, and/or biased towards the most visually conspicuous mergers. Therefore, in this work, we used a second simulation for verification of merger classification methods. This gives us a better understanding of what could happen when applying classifiers trained on simulations to real observations. We used TNG, and more specifically, the combination of TNG300 and TNG100, as our main training sample for two reasons. Firstly, TNG300 has the largest volume within the TNG and Horizon-AGN simulations, which results in the largest sample (together with TNG100, which adds lower-mass galaxies to our training sample). Secondly, we want the secondary simulation, used only for testing, to have a similar or better resolution than the training data. Horizon-AGN has a baryon particle resolution better than TNG300  and similar to TNG100 (even though the DM particle resolution is an order of magnitude lower than TNG100). In this work, we explored and compare in detail how six leading ML classification methods perform on the three datasets (TNG, Horizon, and HSC) for two different tasks:

\begin{enumerate}
    \item Binary classification between major mergers and non-mergers. Definitions of major merger and non-merger can be found in Sect. \ref{subsec.challenge.datasets}.
    \item Multi-class classification into four classes (non-merger, pre-merger, ongoing-merger, post-merger). Definitions of these four classes can be found in Sect. \ref{subsec.challenge.datasets}.
\end{enumerate}

\subsection{Mock images}\label{subsec.data.mock}

We explain here how synthetic images of simulated galaxies from both simulations are processed to generate mock images as if they were observed by HSC. For each mock galaxy image, all stellar particles around the main galaxy were used (within the size of the mock image) to ensure that secondary galaxies that will eventually merge with the main one are visible in the image. Each stellar particle from the simulations contributes its own spectral energy distribution derived from the \cite{Bruzual2003} stellar population synthesis models, depending on its mass, age, and metallicity. The sum of the contribution from all stars passes through the desired filter to create a smoothed 2D projected map \citep{Rodriguez-Gomez2019, Martin2022}. These maps do not include a full radiative transfer treatment and, therefore, do not account for dust. For this work, the simulated images were produced in the \textit{i}-band, with the HSC pixel resolution, and had a physical size of ~$160\times160$ kpc. We chose this size as this is the maximum separation between merging galaxies based on binary merger simulations \citep{Qu2017, Moreno2019}. Then, each image was convolved with the \textit{i}-band PSF, retrieved from the HSC-SSP database. The third step is to add Poisson noise. Lastly, each image was injected into cutouts of real HSC observations. 

For the final step, we need cutouts of the real HSC sky, which should not have bright sources in the centre where the synthetic images will be injected. To ensure this, we constructed a catalogue of low-$z$ and bright sources which we want to avoid using the following criteria: $129^{\circ} \le \mathrm{RA} \le 180^{\circ}$, $-2^{\circ} \le \mathrm{Dec} \le 2^{\circ}$, $z\le 1$, $g_{cModel} \le 26.0$, $r_{cModel} \le 25.6$, $i_{cModel} \le 25.4$, $z_{cModel} \le 24.2$, $y_{cModel} \le 23.4$. This way we still allowed cutouts to contain possible faint sources and higher redshift ($z>1$) background galaxies, as would be the case for real observations. We generate sky cutouts centred on random sky coordinates, keeping only those that do not have any catalogued bright or low-$z$ source within 21{\arcsec} (based on the surface density of the sources to be avoided). After that, we discarded cutouts that, according to mask flags \citep{Bosch2018}, contain bad pixels, saturated pixels, unmasked NaN, and possible missed bright objects. The whole process is illustrated in Fig. \ref{fig.mock_images_process}, from the raw simulated image to the PSF-convolved image, to the addition of Poisson noise, and finally the injection into the real sky from the HSC survey. As stated before, this procedure has the limitation of not including a full radiative transfer treatment and not including the effects of dust. However, given the wavelength used in this study ($i$-band), we are probing the rest-frame optical, except for galaxies at redshift > 0.9. Only for galaxies above this redshift are we probing the rest-frame ultraviolet, where the effect of dust will be of greater importance. Furthermore, \cite{Bottrell2019} show that radiative transfer effects for gas-rich, star-forming galaxies (which will be most affected by dust obscuration) do not have a significant impact on the performance of ML models trained to classify merger galaxies, as the models appear to focus on broad morphological features (such as tidal features) rather than variations due to dust obscuration. Dust effects, therefore, produce only a slight improvement in performance, while realistic instrumental effects, such as PSF resolution and realistic noise and crowding of nearby sources, are much more important.

We split the training set into four redshift bins: $z1\rightarrow[0.1, 0.31)$, $z2\rightarrow[0.31, 0.52)$, $z3\rightarrow[0.52, 0.76)$, $z4\rightarrow[0.76, 1.0)$. The redshift bins were chosen so that each bin has a similar redshift span with a similar number of galaxies. At different redshifts, a physical size of 160 kpc corresponds to a different image size in pixels. For each redshift bin, we used the image size, in pixels, that corresponds to 160 kpc in the midpoint of the $z$ range of the bin, which corresponds to image sizes of 320, 192, 160, and 128 pixels. The final adopted sizes for each method may vary, as some use smaller sizes than the ones provided. Figure \ref{fig.mock_images_examples} shows examples of mock HSC images of simulated galaxies from TNG and HorizonAGN in the four redshift bins. We can clearly see how much more difficult it is to discern features at higher redshifts, at least visually.

\subsection{Training and testing datasets}\label{subsec.challenge.datasets}

For IllustrisTNG, a complete merger history is available through merger trees for each galaxy \citep{Rodriguez2015}, and similarly for Horizon-AGN. We used these trees to construct our samples from both simulations. For the first task (binary classification), we needed a sample of mergers and a sample of non-mergers. For the multi-class task, we needed to subdivide the merger class into three subclasses (pre-mergers, ongoing-mergers, and post-mergers). Following the merger trees, we constructed a sample of mergers, which were selected to be galaxies that had a merger event in the last 0.3 Gyr or will have a merger event in the following 0.8 Gyr. Only major mergers with stellar mass ratios  $>$ 1:4 were included. Furthermore, mergers that are -0.8 to -0.1 Gyr, -0.1 to 0.1 Gyr, and 0.1 to 0.3 Gyr away from the coalescence ($dt=0$ Gyr) were classified as pre-mergers, ongoing-mergers, and post-mergers, respectively.  Figure \ref{fig.merger_sequence} shows examples of simulated galaxies from TNG at different merger stages. A control sample of non-mergers for each simulation consists of galaxies that do not follow the criteria above. These galaxies are much more numerous, so we selected a random sample to roughly match the merger sample size, with the same mass and $z$ criteria. The stellar mass and redshift distributions of the TNG and Horizon-AGN datasets are shown in Fig. \ref{datasets}.

\begin{figure*}
    \centering
    \includegraphics[width=1\textwidth]{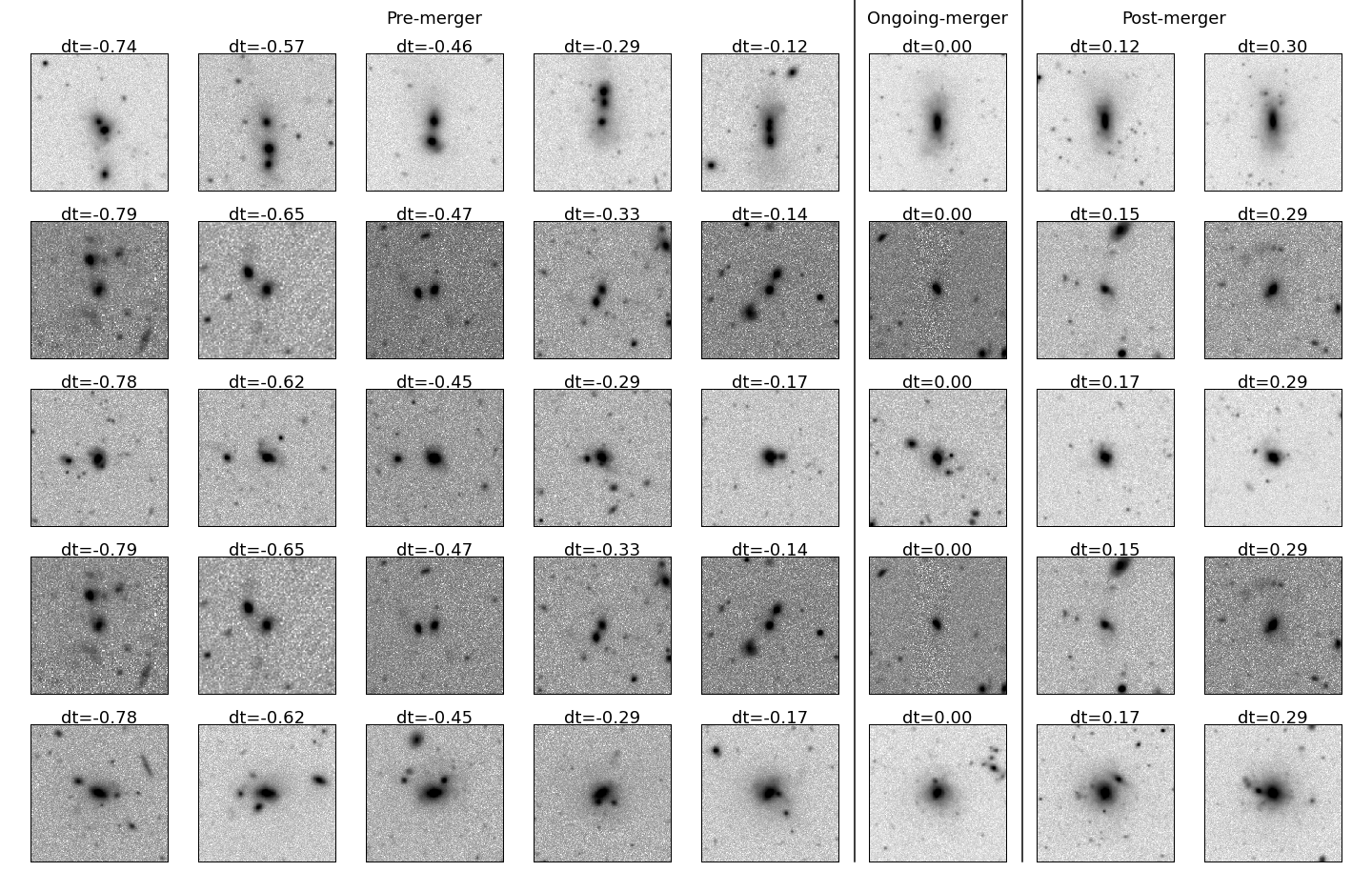}
    \caption{Example mergers from TNG at different merger stages (obtained from the corresponding merger trees in the simulation): pre-mergers ($-0.8<dt<-0.1$ Gyr), ongoing-mergers ($-0.1<dt<0.1$ Gyr), and post-mergers ($0.1<dt<0.3$ Gyr). Each row shows a galaxy along its merger sequence. Images have an approximate physical size of 160 kpc, displayed using an arcsinh inverted grey scale.}\label{fig.merger_sequence}
\end{figure*}

\begin{figure*}
    \centering
    \includegraphics[width=0.49\textwidth]{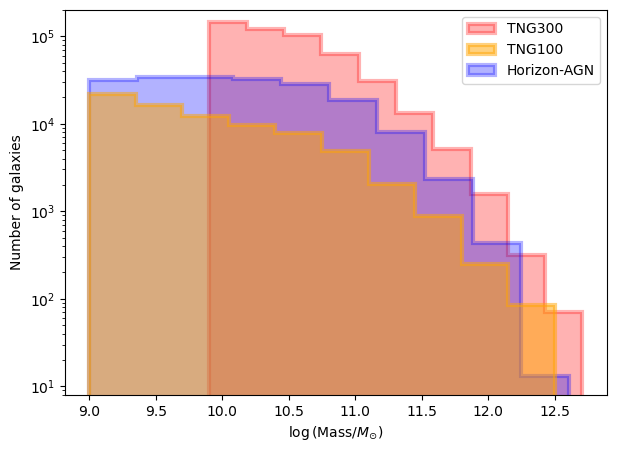}
    \includegraphics[width=0.49\textwidth]{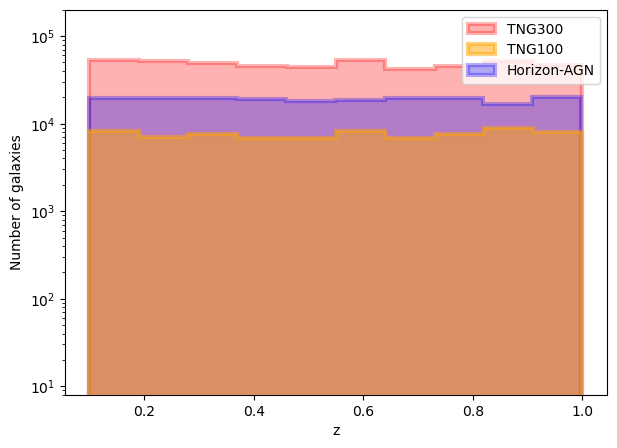}
    \caption{Stellar mass (left) and redshift (right) distributions of the different simulated datasets (red: TNG300; yellow: TNG100; blue Horizon-AGN). By combining TNG100 and TNG300, a broader range of stellar masses is covered. The redshift distributions of the simulation are not designed to follow the observations.}\label{datasets}
\end{figure*}

The IllustrisTNG dataset is split into a training sample (TNG-Training) and a testing sample (TNG-Test). The former comprises 90\% of the whole sample and is used for training the different ML methods. The latter comprises the remaining $10\%$ and was used to measure the performance of the methods on images that have not been seen by the algorithms during training. Galaxies belonging to the same merger tree end up in only one of these splits, which ensures that the test sample cannot be learned by interpolation from the training sample \citep{Eisert2023}. Along with the TNG-test, Horizon-AGN was also used to quantify the performance of the methods. The Horizon-AGN samples have labels (i.e. merger--non-merger for the binary task and non-merger--pre-merger--ongoing-merger--post-merger for the multi-class task) and were only used to evaluate the performance metrics (so not used in training). As seen in Fig. \ref{datasets}, the Horizon-AGN stellar mass distribution does not follow TNG100 (or TNG300, where the lower mass limit is $8\times10^{10}M\odot$). The differences may arise from the different design choices of the simulations. The difference in the stellar mass distributions is not necessarily a problem for training ML models, as long as the training set has enough galaxies that cover the whole stellar mass range of the test set. However, the lower number of galaxies at the low-mass end, compared to the total number of galaxies, may have an impact on the performance of the models for low-mass galaxies, as discussed in Sect. \ref{subsubsec.results.tng.binary} and \ref{subsubsec_results_horizon_binary}.

The HSC sample, constructed to have similar stellar mass and $z$ ranges as the simulations, consists of randomly selected galaxies in GAMA-09. The HSC images (with a physical size of 160 kpc) were split into the four redshift bins and resized according to the image size (in pixels) of each bin to match the simulated image sizes. The HSC sample has no true classification labels, as explained before. So, we included the 2111 galaxies visually classified by \citep{Goulding2018} (see Sec. \ref{subsec.data.hsc}). There are 1243 galaxies classified as clear major mergers, while the other 868 are classified as disturbed--minor mergers. Labels for this visual sample were not shared with the participants. In Table \ref{table_data} we summarise the three samples from IllustrisTNG, Horizon-AGN and HSC. There are more than twice as many objects in TNG than in Horizon-AGN. In both simulations, the merger and non-merger sample sizes are similar to each other by construction. Among the different merger stages, sample sizes of ongoing- and post-mergers are much smaller (due to the shorter timescales) compared to pre-mergers, which could have an impact on the performance of the classifiers in detecting these later merger stages.

\begin{table*}\label{tab.data}
\caption{Total number of galaxies in the training sample (TNG-train) and testing sample (TNG-test, Horizon-AGN, and HSC). We show how many galaxies in each sample are mergers (and more specifically pre-, ongoing-, and post-mergers, as determined from the merger trees) and non-mergers. For the HSC sample, we show the total number of galaxies along with galaxies that were visually classified as merger, non-merger, and disturbed or minor-mergers, as described in Sect. \ref{subsec.challenge.datasets}.}\label{table_data}
    \centering
    % \small
    \begin{tabular}{l|lllllll}
    \hline
    Sample        & Total  & Non-merger    & Merger         & Pre-merger & Ongoing-merger & Post-Merger & Disturbed    \\
    \hline
    TNG-Training  & 504741 & 252473 (50\%) & 252268 (50\%) & 177278 (35\%) & 34574 (7\%) & 43005 (8\%) & --    \\
    TNG-Test      &  56765 &  28280 (50\%) &  28485 (50\%) &  19959 (35\%) &  3908 (7\%) &  4912 (8\%)  & --   \\
    Horizon-AGN   & 196835 &  80687 (41\%) & 116148 (59\%) &  62096 (32\%) & 17597 (9\%) & 36455 (18\%)  & --  \\
    HSC           & 123873 &   1095    &   1243      &      --       &  --         & --  & 868\\
    \hline
    \end{tabular}
\end{table*}

\subsection{Evaluation metrics}\label{subsec.challenge.metrics}

To evaluate the performance of different methods we used several popular metrics for classification problems, including accuracy, precision, recall, F1-score, receiver operating characteristic (ROC) curve, and area under the ROC curve (AUC).

Accuracy is the fraction of correctly classified examples. However, it does not distinguish between the fraction of correctly identified examples from each class. In some cases, correctly identifying a particular class is more important, so accuracy may not be the most appropriate metric. Precision is the fraction of examples from one class correctly identified (reliability for that class), and recall is the fraction of examples from one class that are correctly identified (completeness). Unfortunately, when the classes are difficult to separate (e.g. mergers and non-mergers), it is normally not possible to have both high precision and high recall. This is called the precision-recall trade-off, which means increasing precision will decrease recall and vice-versa. Precision and recall can be calculated for each class as follows,

\begin{equation}
    \text{Precision}=\frac{\text{TP}}{\text{TP}+\text{FP}},
\end{equation}

\noindent and

\begin{equation}\label{eq_recall}
    \text{Recall}=\frac{\text{TP}}{\text{TP}+\text{FN}}.
\end{equation}

For mergers, TP (true positives) is the number of mergers correctly classified as mergers. FP (false positives) is the number of non-mergers incorrectly classified as mergers. FN (false negatives) is the number of mergers classified as non-mergers. The $F_1$ score is the harmonic mean of the precision and the recall,

\begin{equation}
    F_1=\frac{2}{\text{Recall}^{-1} + \text{Precision}^{-1}}.
\end{equation}

The ROC curve shows the performance of a classifier by plotting True Positive Rate (TPR; a synonym for recall) versus false positive rate (FPR) at different classification thresholds, where

\begin{equation}
    \text{TPR}=\frac{\text{TP}}{\text{TP}+\text{FN}},
\end{equation}

\noindent and

\begin{equation}
    \text{FPR}=\frac{\text{FP}}{\text{FP}+\text{TN}}.
\end{equation}

A perfect classifier would yield a point in the upper left corner with a coordinate (0,1), while a random classifier would show a diagonal line. By lowering the classification threshold, more items are classified as positive, thus increasing both FP and TP. Lastly, AUC corresponds to the area underneath the ROC curve, and measures how well predictions are ranked, rather than their absolute values. AUC quantifies the model's performance irrespective of the classification threshold, with ranges from 0 (all predictions are incorrect) to 1 (all predictions are correct).

\section{Machine learning-based merger detection methods}\label{sec.methods}

We explored six different methods from two categories: traditional feature-based ML and image-based DL methods. The first method employs a RF algorithm, while the rest use CNNs. In this section, we briefly describe the main characteristics of each method, for instance the structure; the number of images used for training, validation, and testing; image pre-processing (if any); whether different redshift bins are analysed separately or combined; whether the multi-class classification is provided in addition to the binary classification; and how thresholds for different classes are chosen. In Table~\ref{tab.summary.methods} (in Appendix \ref{appendix_summary_table}) we summarise the methods explored in this study, highlighting the main differences and similarities (for more details on each method, see the list of references).

\subsection{Method-1 (RF)}\label{subsec.methods.1}

This method performs binary and multi-class classification tasks using the RF algorithm \citep{Ho1995}, which is an ensemble learning method that fits multiple decision tree classifiers to various subsamples of the dataset. The final classification for a particular example is then obtained by averaging the classifications from all the individual trees. For data pre-processing, we first performed source deblending to separate overlapping but different sources for each image in all datasets, in order to isolate the galaxy of interest and remove unwanted or contaminating sources from the calculation of the morphological diagnostics. This procedure is described in \cite{Guzman-Ortega2023}. Secondly, we run \texttt{statmorph} \citep{Rodriguez-Gomez2019}, which is a code for calculating non-parametric morphological diagnostics of galaxy images (such as the Gini-$M_{20}$, asymmetry and concentration parameters), as well as fitting 2D S\'ersic profiles. The resulting statistics from applying \texttt{statmorph} to each galaxy image were subsequently employed as model features for the classifier. In particular, we used the measurements of 32 parameters (including Gini, concentration and asymmetry; for a complete list, please check \citealp{Rodriguez-Gomez2019}) as features, ensuring that only reliable quantifications are used.

For the binary task, we used the extracted features from TNG-training. Using the scikit-learn library \citep{Pedregosa2011}, we performed a 5-fold cross-validation strategy together with 200 iterations of a randomised search for tuning the RF hyper-parameters of $n\_\text{estimators}$ (the number of trees in the forest), $\text{max}\_\text{depth}$ (the maximum depth of the tree), $\text{min}\_\text{samples}\_\text{split}$ (the minimum number of samples required to split an internal node), and $\text{min}\_\text{samples}\_\text{leaf}$ (the minimum number of samples required to be at a leaf node). We then obtained a set of these parameters that maximise the accuracy score for all combinations in the random search and used them to refit the RF classifier on the corresponding training set for each redshift interval. The resulting model was applied to the corresponding testing sets to obtain merger predictions. The multi-class task was carried out in a similar fashion, except that we employed the balanced RF algorithm from the \textit{imblearn} library \citep{Nogueira2017}, which uses random undersampling to deal with class imbalance.

\subsection{Method-2 (Swin) }\label{subsec.methods.2}

This is a DL method that uses a Swin Transformer architecture \citep{Liu2021} pre-trained on ImageNet-1K\footnote{Dataset containing a collection of 1.2 million labelled images with one thousand object categories \protect\citep{ImageNet1k}, from animals to everyday objects.} data with an additional fully connected layer with $256$ neurons and an output layer of four neurons, each for one of the four classes in the multi-class task (pre-merger, ongoing-merger, post-merger, and non-merger). In terms of data pre-processing, first, the mock images were cropped to $112\times112$ pixels and linearly scaled between 1 and 0. Each image was then stacked with itself to form a 3-channel image (to be compatible with the expected input of the architecture) and resized to $224\times224$ pixels using the nearest neighbour interpolation).

For the binary classification task, we fixed the parameters of the Swin Transformer during training and so only the fully connected layer and the output layer are trained with the TNG-training data. Data augmentation was performed during training with each image randomly rotated by multiples of $90^\circ$, randomly flipped horizontally, and randomly flipped vertically. TNG-training was split into training and validation datasets (80-20 split), ensuring that members of the same merger tree are only found in the training or validation set and not both. The output of the network is a 4-element vector. Each element represents the probability assigned to that class, such that the sum of all four elements is 1. For the binary task, an image was classified as a merger if the probability of any of the merger classes (pre-merger, ongoing-merger, or post-merger) is higher than the probability of being a non-merger. For the multi-class task, the same trained network as for the binary task was used, and an image will have the classification of the class with the highest probability.

\subsection{Method-3 (Zoobot) }\label{subsec.methods.4}

Zoobot \citep{Walmsley2023} is a Python package for measuring the detailed appearance of galaxies using DL. Zoobot includes CNN and vision transformer models pre-trained on the responses of Galaxy Zoo volunteers to (real) images \citep{Willett2013, Willett2017, Simmons2017, Walmsley2022a}. These models are designed to be adapted to new tasks and surveys using minimal new labels. Here, a pre-trained Zoobot CNN model\footnote{EfficientnetB0, pre-trained on GZ Evo as described in \cite{Walmsley2022c}. This earlier GZ Evo version did not include classifications from HSC images, i.e. GZ Cosmic Dawn, which was not yet complete at the time of writing.} is adapted to perform the Merger Challenge tasks. For classifying simulated images, we add a custom head and loss designed to jointly predict the answers to each task. Specifically, our head is a single dense layer with five outputs, as follows,

\begin{itemize}
    \item Two outputs classify whether a galaxy is a merger or non-merger.
    \item Three outputs classify the subclasses (pre-merger, ongoing-merger, and post-merger).
\end{itemize}

The outputs use a cross-entropy loss, and no activation functions are applied. Each component of the loss is weighted to be roughly equal\footnote{We applied weightings of [1, 1, 10, 1, 3] to each group of outputs (with respect to the list above).} and only applied when the task is relevant (e.g. there is no additional loss for predicting merger subclass wrong when the galaxy is not a merger).

Regarding data pre-processing, all images were cropped to a smaller physical size of 100 kpc, with an arcsinh scaling. Hyper-parameters (batch size, image physical size, and AdamW weight decay) were optimised based on a grid search training only on the lowest-$z$ subset (for speed). Experiments showed that the best performance was obtained when all model parameters were set as trainable, not just the head or final convolutional layers. Initialising from the pre-trained Zoobot model significantly outperformed initialising an otherwise identical model from random.

\subsection{Method-4 (CNN1)}\label{subsec.methods.3}

This method uses the CNN architecture described in \cite{Bickley2021}, which has been used for post-merger versus non-merger classification. We trained four networks, one for each redshift bin. The networks have four convolutional layers with 32, 64, 128, and 128 filters, respectively, all of size $7\times7$, followed by two dense layers with 512 and 128 neurons and finally by a 1-neuron dense layer. Each convolutional layer is followed by a max-pooling layer and a dropout layer. Dropout is also applied to the dense layers. For data pre-processing, the images were first cropped to $120\times120$, $96\times96$, $80\times80$ and $64\times64$, for $z$-bin 1, 2, 3 and 4, respectively. This corresponds to approximately a physical size of 80 kpc for all redshift bins. For the first redshift bin, the size corresponds to 60 kpc. This was done to reduce the size of the network, memory, and time needed for training. We used arcsinh scaling, along with clipping to maximise the contrast in the central region. Finally, we normalise the images between 0 and 1.

For the binary task, we trained each redshift bin separately. In each bin, TNG-training was split into training and validation datasets (90-10 split), making sure that the same merger histories were not split between the sets. Each network was optimised with the ADADELTA optimiser \citep{Zeiler2012}. Data augmentation was used during training, including random rotation between 0 and 90$^{\circ}$, random horizontal and vertical flip, and random zoom with a factor between 0.7 and 1.3. The hyper-parameters (such as batch size, learning rate, and optimiser) of the network were fine-tuned to find the best performance on the first redshift bin. The same set-up is used for the other $z$ bins. We used early stopping in each network to ensure that there is no over-fitting. After finding the best models, we applied them to TNG-test, Horizon-AGN, and HSC. The output of the network is a value between 0 and 1. We used a threshold of 0.5 to separate non-mergers and mergers. For the multi-class task, we first balanced the datasets by performing data augmentation for the ongoing-mergers and post-mergers classes. The same networks and hyper-parameters as for the binary task were used, but the last layer was replaced by a softmax function with four neurons (one for each class). The classification corresponded to the class with the highest probability (given by the corresponding neurons in the last layer).

\subsection{Method-5 (CNN2)}\label{subsec.methods.5}

This method used a CNN architecture (Chudy et al. in prep.) with four convolutional layers, each consisting of 128 filters with sizes  $13\times13$,  $11\times11$, and $11\times1$, respectively.
After each convolutional layer, the ReLu activation function, batch normalisation, dropout (of 20\%), and max pool layer are applied. After convolving, the output is flattened before being put forward to the next two fully connected layers of 512 and 128 neurons, respectively. After each dense layer, activation, batch normalisation, and dropout (of 20\%) are performed. The output of a network is a single neuron activated by a softmax function that provides values representing the probability of being a merger. The loss of the network is determined using binary cross entropy. The CNN has 8186113 total trainable parameters. Regarding data pre-processing, all images were first resized to 128x128 pixels and normalised between 0 and 1. Images were randomly divided using a 75:15:10 ratio into three sets: the training set used to fit the parameters, the validation set used to evaluate a model while tuning the model’s hyper-parameters, and finally, the test set used for an unbiased evaluation of the final model. 

For the binary task, the network was trained on all TNG-training data from the four redshift bins. It was optimised with the ADAM algorithm with a learning rate $\alpha=3\times10^4$ and trained for 112 epochs with the set of tuned hyper-parameters for the epoch that provided the highest value of validation accuracy used for classification. The threshold for classification is set to 0.51, at which TPR = True Negative Rate (TNR). This method has a similar architecture to Method-4 (CNN1), but differs in the number and size of filters, resulting in a higher number of trainable parameters. Another key difference is the scaling used for the images (linear scaling in this case). While for Method-4 (CNN1) the images are cropped to a physical size of 80 kpc to focus more on the central region, for this method, the original size of 160 kpc is used. Lastly, for Method-4 (CNN1), we trained four networks, one for each redshift bin, while for this method, we trained a single network for all redshift bins. This means that the higher the redshift bin, the more compressed the images are, which could result in loss of information.

\subsection{Method-6 (CNN3)}\label{subsec.methods.6}

This method uses a CNN architecture \citep{Walmsley2019}, consisting of three convolutional layers. The first convolutional layer has 32 filters with size $3\times3$, the second layer has 38 filters with size   $3\times3$, and the last layer has 64 filters of size  $2\times2$. Each convolutional layer is followed by a pooling layer. The convolutional part is then followed by a 64 neurons dense layer and an output 1-neuron layer, activated by a sigmoid function, providing an output between 0 and 1 for each image. The loss of the network consists of binary cross entropy. The network has a total of  616,497 trainable parameters. Regarding data pre-processing, all images were resized to an image size of $100\times100$ pixels. To normalise the images we applied the function transform = AsinhStretch(0.1) + PercentileInterval(97), where AsinhStrtch(0.1) performs the following operation on an image x: asinh(x/0.1)/asinh(1/0.1), and PercentileInterval(97), looking at the cumulative pixel distribution, sets the lowest 1.5\% to 0 and the highest 1.5\% to 1, while the rest 97\% is normalise between 0 and 1.

For the binary task, the network was trained on TNG-training data from the four redshift bins. The training data was split into training and validation with a 90:10 split, ensuring no merger history was split between the sets. The network was optimised with the ADAM algorithm with a learning rate $\alpha=0.001$. Data augmentation was used during training, including random rotation between -45$^\circ$ and 45$^\circ$, random vertical flip, random horizontal flip, random translations ($\pm5\%$), and random zoom (with a factor between 0.75 and 1.3). The threshold for classification is set to 0.51, which maximises TPR and minimises TNR. This method has a similar architecture to Method-4 (CNN1) and Method-5 (CNN2). However, the architecture in this case is simpler, with a smaller number of layers and trainable parameters. Similar to Method-5 (CNN2), a single network was trained for all $z$ bins combined. The images have a physical size of 160 kpc, which is twice as much as used in Method-4 (CNN1), but the same as in Method-5 (CNN2), albeit with smaller image size (in pixels) and potentially more loss of information in the highest $z$ bins. It is interesting to investigate these three CNNs to determine whether the differences, such as the number of layers, image scaling, or use of data augmentation, can have a significant impact.

\section{Results}\label{sec.results}

\subsection{TNG}\label{subsec.results.tng}

In this section, we show the performance of the six different methods on the TNG-test dataset using a variety of metrics for the binary task and the multi-class task. First, we check the overall performance by combining all redshift bins. Then, we examine in detail how performance changes as a function of redshift (for both tasks) and stellar mass (only for the binary task). 

\begin{figure}
    \centering
    \includegraphics[width=0.48\textwidth]{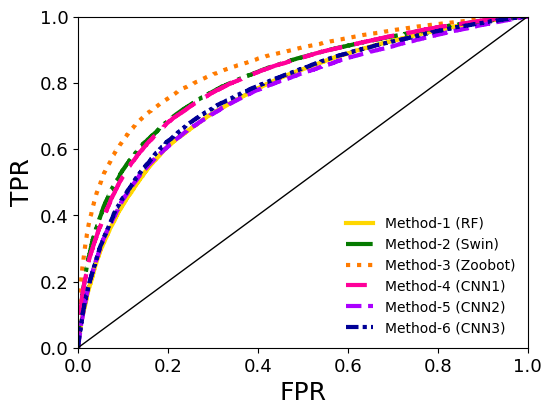}
    \caption{ROC for the TNG test set. The ROC curves show the overall performance of each method independently of the chosen classification threshold. The farther the curve is from the 1:1 line (which represents a random classifier) or the greater the area under the curve, the better the model. Method-3 (Zoobot) shows the best performance in terms of ROC.}\label{plot_roc_TNG}
\end{figure}

\begin{figure*}
    \centering
    \includegraphics[width=0.9\textwidth]{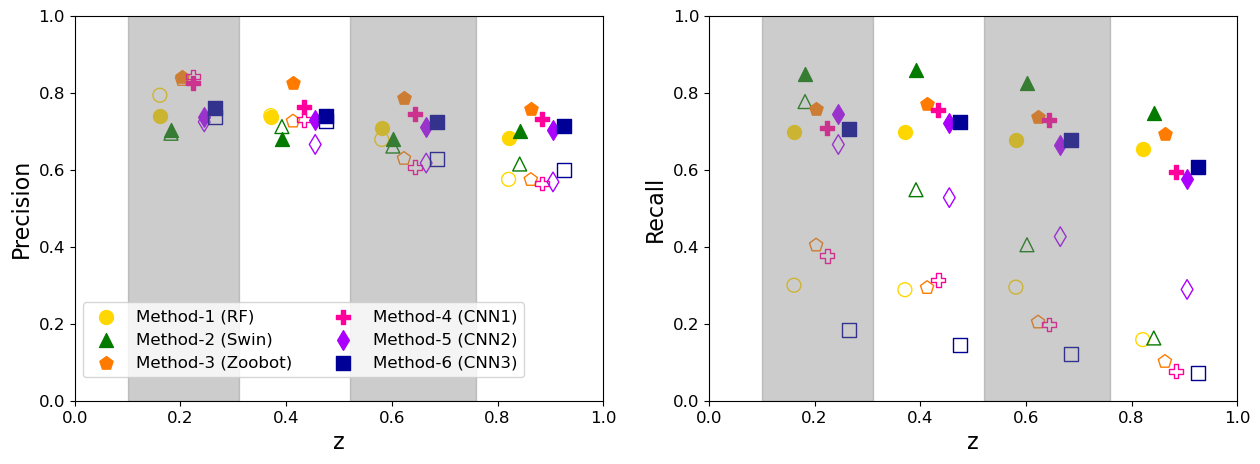}
    \caption{Precision (left) and recall (right) of the merger class as a function of redshift. The filled symbols correspond to the performance of the methods on the TNG dataset and the empty symbols correspond to those on the Horizon-AGN dataset. There is a slight downward trend in precision and a more significant drop in recall with increasing redshift for TNG. These trends are stronger for Horizon-AGN. While precision and recall are both relatively high for TNG, for Horizon-AGN recall drops much more than precision. All methods were trained on TNG and then applied to the Horizon-AGN dataset. }\label{fig_prec_recall_TNG}
\end{figure*}

\begin{figure*}
    \centering
    \includegraphics[width=0.9\textwidth]{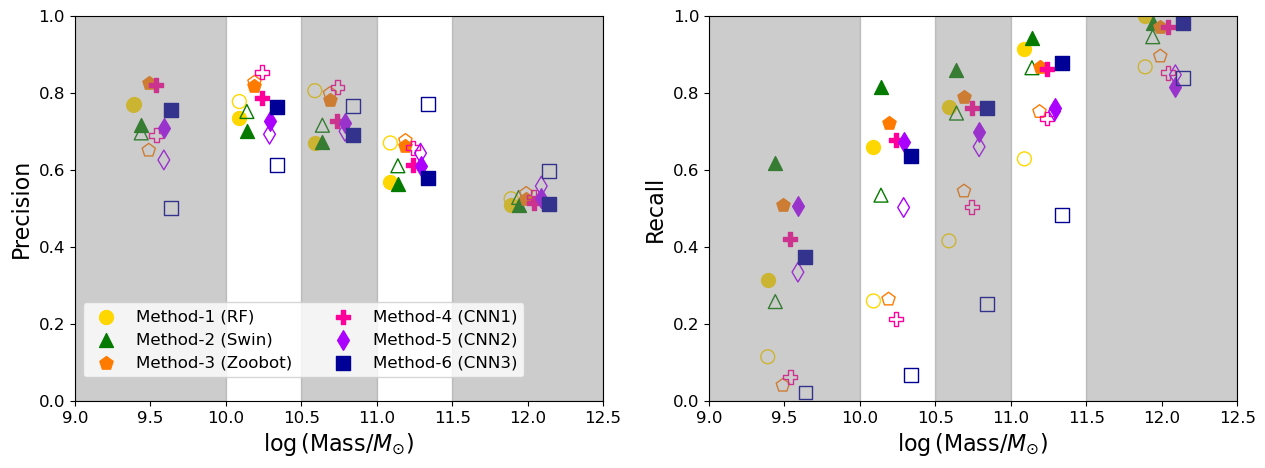}
    \caption{Precision (left) and recall (right) of the merger class as a function of stellar mass for each method. The filled symbols show the metrics for the TNG dataset and the empty symbols for the Horizon-AGN dataset. For most methods and in both simulations, precision remains constant with mass, but then decreases with increasing mass at $M_{*} > 10^{11}M_{\odot}$. There is a sharp downward trend in recall with decreasing mass for both datasets. }\label{fig_prec_recall_TNG_mass}
\end{figure*}

\subsubsection{TNG: binary classification}\label{subsubsec.results.tng.binary}

\begin{table}
\caption{Performance metrics as percentages (accuracy, precision, recall, and F1-score for the merger class and AUC) of the different methods on the TNG-test set for the binary classification task. The best performance in each metric is highlighted in bold.}\label{tab.tng_all_redshifts}
    \centering
    % \small
    \begin{tabular}{l|ccccc}
    \hline
              & \multicolumn{4}{c}{All redshift bins combined}     \\
    \hline
    Methods   & Acc.  & P     & R     & F1    & AUC \\
    \hline
    Method-1 (RF)  & 70.5  & 71.6  & 68.1  & 69.8  & 77.0\\
    Method-2 (Swin) & 72.6  & 69.1  & \bf{81.9}  & 74.9  & 81.2\\
    Method-3 (Zoobot) & \bf{77.7}  & \bf{80.1}  & 73.8  & \bf{76.8}  & \bf{85.1}\\
    Method-4 (CNN1) & 74.1  & 76.6  & 69.5  & 72.8  & 80.9\\
    Method-5 (CNN2) & 70.6  & 72.0  & 67.2  & 69.5  & 76.5\\
    Method-6 (CNN3) & 71.5  & 73.3  & 67.6  & 70.4  & 76.9\\
    \hline
    \end{tabular}
\end{table}

In Table \ref{tab.tng_all_redshifts} we summarise the performance of the different methods on the whole TNG-test sample over the entire redshift range explored, using various metrics including accuracy, precision, recall, F1-score and AUC. We note that precision and recall are only calculated for the merger class. Overall, all methods show a similar performance on TNG-test, with a maximum difference of 12\% in precision (ranging from $\sim$69\% to 81\%) and 15\% in recall (ranging from $\sim$67\% to 82\%). However, these methods rely on choosing a threshold on the output probability to classify mergers and non-mergers. A common value of this threshold is 0.5, but different thresholds will yield different precision and recall. Another common choice of threshold is the value at which TPR = TNR. In general, one can increase precision at the cost of reducing recall, and vice versa. Metrics that are independent of the threshold choice are the ROC and AUC, which give an overall view of the performance of the model. Fig. \ref{plot_roc_TNG} shows the ROC for all six methods. Method-3 (Zoobot) shows the highest AUC at just over 85\%, while also having the highest accuracy (78\%), precision (80\%) and F1 score (77\%). Method-2 (Swin) has the highest recall (82\%) and the second highest F1-score and AUC, but its precision (69\%) is the lowest. Method-4, -5, and -6, which use similar CNNs (with small differences in the exact architecture, image pre-processing and augmentation), achieve similar performance in all metrics, varying by only a few per cent. The performance of Method-1, which is the only traditional ML method in this study, is actually quite similar to the worst-performing DL-based methods.

We also evaluate the performance of the models in each redshift bin, as shown in Table \ref{tab.tng_redshift_bins}, in Appendix \ref{appendix.results.redshift}. Precision and recall are plotted as a function of redshift in Fig. \ref{fig_prec_recall_TNG} as filled symbols. Method-3 (Zoobot) has the highest precision at all redshifts, varying between 76\% in $z$-bin4 and 84\% in $z$-bin1. Method-3 also has the highest F1-score in all redshift bins, except in $z$-bin 4 where it is overtaken slightly by Method-2. On the other hand, Method-2 (Swin) has the highest recall at all redshifts, but again its precision is the lowest (except in $z$-bin4 where the traditional ML method, i.e. Method-1, has the lowest precision). There is a mild downward trend with increasing $z$ in both precision and recall for most methods, with an average drop from the lowest to the highest redshift bin of $\sim$5\% and $\sim$9\% in precision and recall, respectively.  

Figure \ref{fig_prec_recall_TNG_mass} shows how precision and recall change with stellar mass (again plotted as filled symbols for the results on the TNG-test). Precision remains more or less constant at $M_{*}<10^{11}M_{\odot}$, and then decreases towards more massive galaxies. However, recall shows a steep drop as stellar mass decreases, with a difference of 30\% to 60\% (depending on the method) from $M_*\sim10^{12}M_{\odot}$ to $M_*\sim10^{9.5}M_{\odot}$. Therefore, it seems that stellar mass has a bigger impact on the performance of the classifiers than redshift, which may be caused by the specific stellar mass distribution of the simulated galaxies in the training data. In comparison, the redshift distribution of the simulated galaxies is much flatter which could lead to less variation in performance as a function of redshift. In terms of precision, the best results are obtained for galaxies with $M_{*}<10^{11}M_{\odot}$. However,  the average recall at $M_{*}<10^{10}M_{\odot}$ is only $\sim$50\%. Galaxies with $10^{10}<M_{*}<10^{11}M_{\odot}$ (where most galaxies are located) show the best balance in terms of precision and recall. The low precision and high recall at $M_{*}>10^{11}M_{\odot}$ can be explained first, by the small number of galaxies in that mass range and secondly, by the small fraction of non-mergers compared to mergers (two times less at $10^{11}<M_{*}<10^{11.5}M_{\odot}$ and five times less at $M_{*}>10^{11.5}M_{\odot}$), which may result in the models learning to predict the most massive galaxies as mergers. In future work, we could try to balance better the number of galaxies in different stellar mass ranges in the training data.

\begin{figure*}
    \centering
    \includegraphics[width=0.4\textwidth]{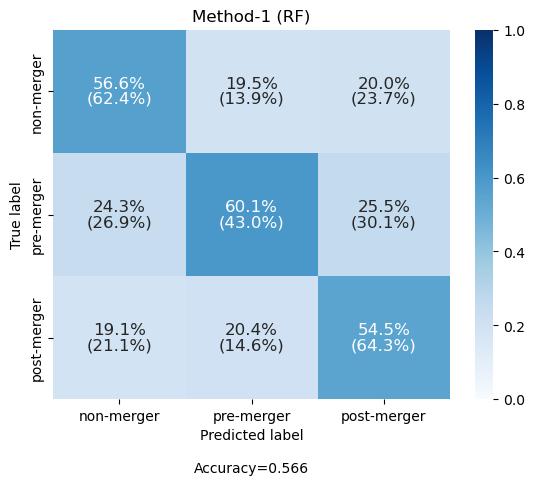}
    \includegraphics[width=0.4\textwidth]{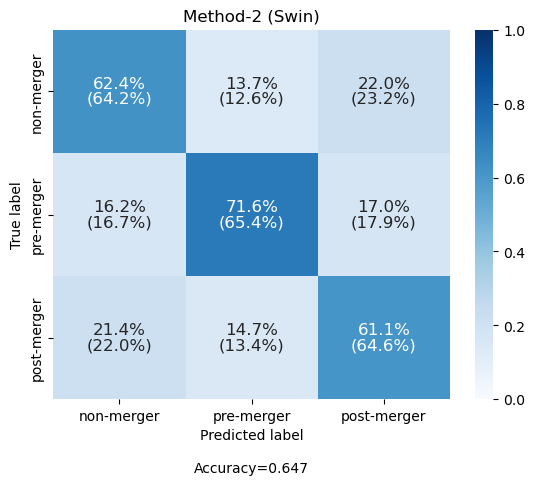}
    \includegraphics[width=0.4\textwidth]{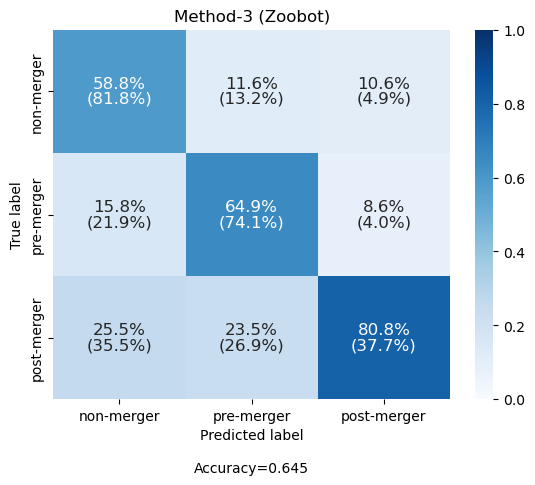}
    \includegraphics[width=0.4\textwidth]{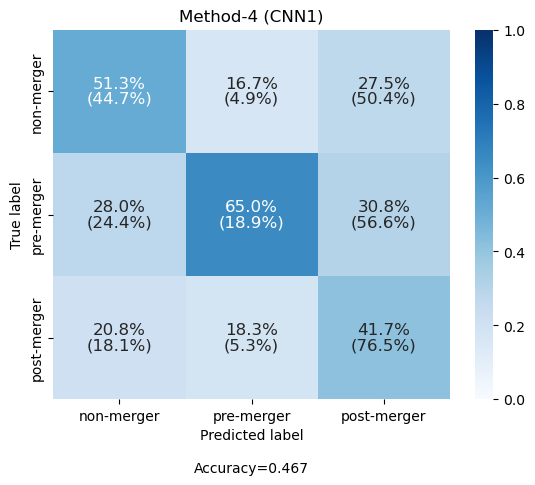}
    \caption{Confusion matrices for Method-1 (top left), Method-2 (top right), Method-3 (bottom left), and Method-4 (bottom right) for the multi-class classification task on TNG. The data from all four redshift bins are combined. The post-merger class includes the ongoing-mergers. The confusion matrices are normalised vertically, and therefore the diagonal elements represent the precision of each class. The recall of each class is shown in brackets.}\label{fig.tng.confusion}
\end{figure*}

\subsubsection{TNG: multi-class classification}\label{subsubsec.results.tng.multi}

Here we present the results obtained for the multi-class classification task. Of the six methods, only four are trained for this task (Method-1 (RF), Method-2 (Swin), Method-3 (Zoobot), and Method-4 (CNN1)). This task aims to predict the four classes corresponding to non-merger, pre-merger, ongoing-merger, and post-merger. All methods show great confusion between ongoing-mergers and post-mergers, probably due to the relatively low number of galaxies in these two classes (corresponding to $7\%$ and $8\%$ of the training set for ongoing- and post-mergers, respectively). Additionally, the morphologies in these two classes tend to be similar making them harder to separate. Therefore, we show in this section the results of combining these two classes into a final combined post-merger class. The results for the 4-class classification are shown in Appendix \ref{appendix.4classes} instead. 

Figure \ref{fig.tng.confusion} shows the confusion matrices for the four methods for all redshift bins combined. The matrices are normalised to show the precision of each class (with recall values shown in brackets). The overall performance is worse than for the binary task, as shown by the lower precision and recall values. An ideal classifier will show values close to 100 in the diagonal and 0 outside. However, all methods show high percentages of misclassifications. For most methods (Method-1, -2, and -4) the easiest class to identify seems to be pre-mergers, with precision of $60\%$, $72\%$, and $65\%$, respectively,  but recall is very low ($42\%$, $65\%$, and $18\%$, respectively). Method-3 (Zoobot) has the highest precision for post-mergers ($81\%$), while for pre-mergers, the precision is $65\%$, comparable to the other methods. However, the recall of Method-3 for post-mergers and pre-mergers is $38\%$ and $74\%$, respectively. 

In Table \ref{tab.tng_multi_redshift_bins} in Appendix \ref{appendix.results.redshift} we show the precision and recall for the pre-mergers and post-mergers as a function of $z$, for each method. The performance of all methods decreases with increasing $z$ ($\sim$10\% and 11\% decrease for precision and recall, respectively), more than for the binary classification task, demonstrating the greater difficulty in distinguishing merger stages with increasing $z$. Some studies in the literature show better performance \citep{Bottrell2019, Ferreira2020}, however, the comparison is not straightforward (as explained in the introduction), as different studies have different definitions of the merger and non-merger classes, or may use better-quality data. In particular, \cite{Ferreira2020} focuses on mock images to mimic data from the Hubble Space Telescope, which may be an indication that deeper data and better spatial resolution can improve performance in distinguishing the different merger stages (as one would naturally expect). Additionally, increasing the relative fraction of the post-mergers compared to the pre-mergers in the training data could also lead to a better performance in merger stage classification.

\subsection{Horizon-AGN}\label{subsec.results.horizon}

This section presents the results of applying the models trained on TNG-training to Horizon-AGN. The use of the second simulation allows us to quantitatively assess how the performance of the various classifiers changes when transferred to a different dataset. This exercise is useful as it gives us an idea of what may happen to the performance of the classifiers when applied to real observations (for which we have no ground-truth labels).

\subsubsection{Horizon-AGN: binary classification}\label{subsubsec_results_horizon_binary}

In Table \ref{tab.horizon_all_redshifts} we summarise the performance of the six methods for all redshift bins combined. The precision for the merger class in Horizon-AGN ($\sim$70\%) does not decrease very much compared to the results on TNG. Method-3 (Zoobot) still achieves the best precision at 72\%, which is 8\% lower than its performance on TNG. Method-1 (RF) and Method-4 (CNN1) are both close second to Zoobot, achieving a precision level of over 71\%. However, we see a much more significant drop in recall relative to TNG. The best-performing method for Horizon-AGN in this metric is Method-5 (CNN2) with a recall of 47\%, which is a 35\% drop compared to the best performance in TNG. The worst-performing method in recall for Horizon-AGN is Method-6 (CNN3) with just over 12\%. Fig. \ref{plot_roc_Horizon} shows the ROC curves for all the models which are fairly similar to each other (as seen also by the AUC values in Table \ref{tab.horizon_all_redshifts}).

\begin{table}
\caption{Performance metrics as percentages (accuracy, precision, recall, and F1-score for the merger class, and AUC) for the different methods (trained on TNG) applied to the Horizon-AGN set, for the binary classification task. The best performance in each metric is highlighted in bold.}\label{tab.horizon_all_redshifts}
    \centering
    % \small
    \begin{tabular}{l|ccccc}
    \hline
              & \multicolumn{4}{c}{All redshift bins combined}     \\
    \hline
    Methods   & Acc.  & P     & R     & F1    & AUC \\
    \hline
    Method-1 (RF) & 57.9  & 71.2  & 26.5  & 38.7  &  67.1 \\
    Method-2 (Swin) & \bf{62.5}  & 68.4  & 46.6  & \bf{55.4}  & \bf{67.2}  \\
    Method-3 (Zoobot) & 57.6  & \bf{72.0}  & 24.8  & 36.9  & 63.1  \\
    Method-4 (CNN1) & 57.1  & 71.7  & 23.5  & 35.4  & 66.0  \\
    Method-5 (CNN2) & 61.2  & 65.4  & \bf{47.3}  & 54.9  &  64.1 \\
    Method-6 (CNN3) & 53.2  & 67.5  & 12.4  & 21.0  & 54.9  \\
    \hline
    \end{tabular}
\end{table}

\begin{figure}
    \centering
    \includegraphics[width=0.48\textwidth]{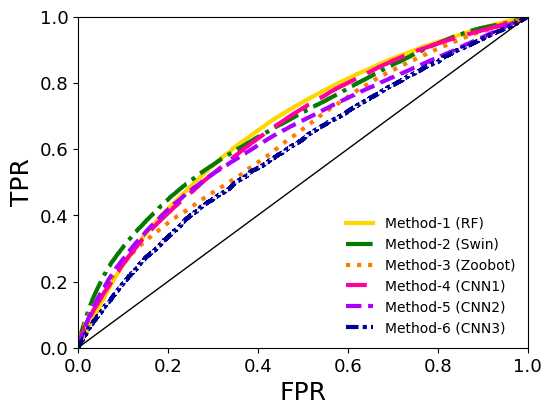}
    \caption{ROC for the different methods (trained on TNG) applied to Horizon-AGN set. Method-1 (RF) and Method-2 (Swin)  show the best performance in terms of ROC. However, the differences with the other methods are small.}\label{plot_roc_Horizon}
\end{figure}

\begin{figure*}
    \centering
    \includegraphics[width=0.4\textwidth]{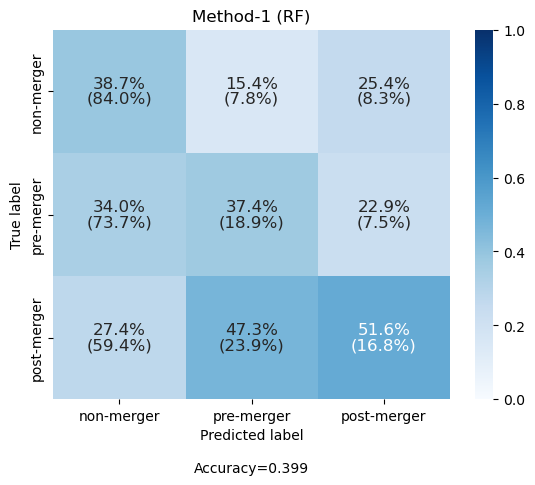}
    \includegraphics[width=0.4\textwidth]{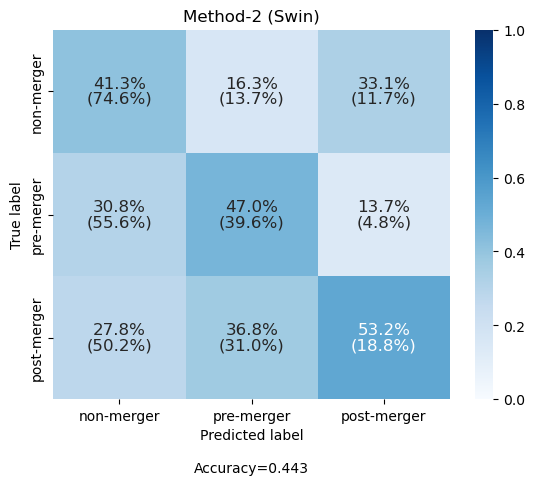}
    \includegraphics[width=0.4\textwidth]{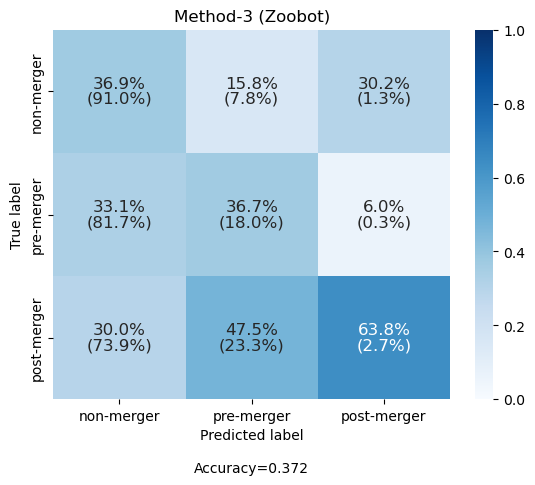}
    \includegraphics[width=0.4\textwidth]{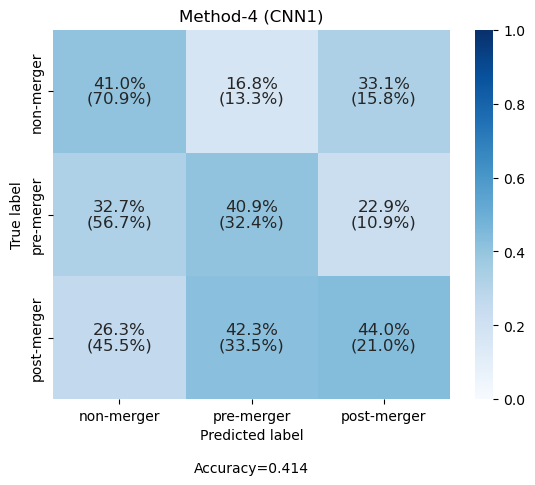}
    \caption{Confusion matrices for Method-1 (top left), Method-2 (top right), Method-3 (bottom left), and Method-4 (bottom right) for the multi-class classification task on Horizon-AGN (from methods trained on TNG). The data from all four redshift bins are combined. The post-merger class includes the ongoing-mergers. The confusion matrices are normalised vertically, and therefore the diagonal represents the precision of each class. The recall is shown in brackets.}\label{fig.confusion.3class.horizon}
\end{figure*}

In Table \ref{tab.horizon_redshift_bins} in Appendix \ref{appendix.results.redshift} we summarise the performance of the methods for each redshift bin, and in Fig. \ref{fig_prec_recall_TNG} we show (in open symbols) precision and recall as a function of $z$. While there is a decrease in the overall precision with respect to TNG, in the first redshift bin, the precisions for both datasets are very similar. Only as $z$ increases does the difference in precision between TNG and Horizon increase. While in TNG-test the drop in precision with increasing $z$ is $<10$\%, in Horizon-AGN precision drops by $\sim$10-30\%, depending on the method. The recall for all methods decreases more sharply as $z$ increases. Only Method-2 (Swin) and Method-5 (CNN3) have comparable recalls to TNG in the first redshift bin, but the difference rapidly increases with $z$. The rest of the methods have recalls $<40$\% already in the first redshift bin. In Fig. \ref{fig_prec_recall_TNG_mass} we show (in open symbols) how precision and recall vary with stellar mass. Precision remains more or less the same for all mass bins as for TNG, except for Method-6 (CNN), which has lower precision for $M_{*}<10^{11}M_{\odot}$ than it had on TNG. Similar to TNG, recall drops rapidly towards lower mass, but in this case, the drop in recall from the highest to the lowest stellar mass bin is even bigger. The discrepancy between the results obtained on TNG and Horizon-AGN may arise from a number of factors (or combinations thereof), such as the different underlying galaxy physics implemented, the difference in the effective resolution, and in how galaxies are identified and linked through time. We expect that the two latter factors play a smaller role as, on one hand,  both simulations use stellar particles to find galaxies (and both simulations have comparable baryonic matter resolution), and on the other hand, they use the same methodology to define mergers through the mergers trees. The different sub-grid physics may result in different galaxy populations (see Appendix \ref{sec.appensix.comparison.data} for a comparison of the galaxy populations), which could have a bigger impact on the difference in performance on the two simulations. All these dissimilarities together could lead to differences in the mock images of mergers produced by each simulation, with our results suggesting that Horizon-AGN produces a type of mock images of mergers not found on TNG.

\subsubsection{Horizon-AGN: Multi-class classification}\label{subsubsec.results.horizon.multi}

In this section, we present the results of applying the four models trained for the multi-class task on the Horizon-AGN dataset. As in Sect. \ref{subsubsec.results.tng.multi}, we show the results after combining the ongoing-mergers and post-mergers classes into the post-merger class. Fig. \ref{fig.confusion.3class.horizon} shows the four confusion matrices for all the Horizon-AGN redshift bins combined. None of the models performs well on this dataset. This could be at least partly due to the fact that the baseline performance (on the training dataset, as seen in Fig. \ref{fig.tng.confusion}) is not high enough to apply to a different domain, in which the performance is expected to drop (as seen in the previous section). 

\begin{figure}
    \centering
    \includegraphics[width=0.48\textwidth]{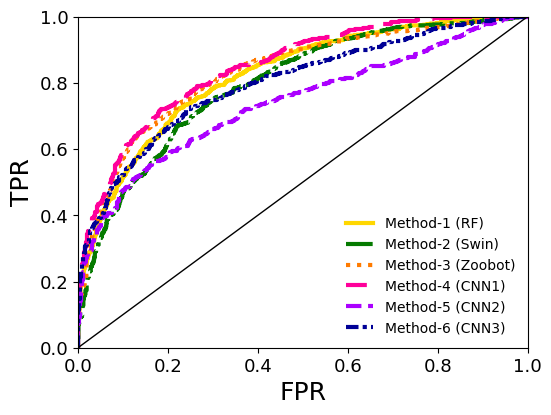}
    \caption{ROC for different methods (trained on TNG) applied to HSC set. Method-4 (CNN1) shows the best performance in terms of ROC.}\label{plot_roc_HSC}
\end{figure}

\begin{figure*}
    \centering
    \includegraphics[width=0.9\textwidth]{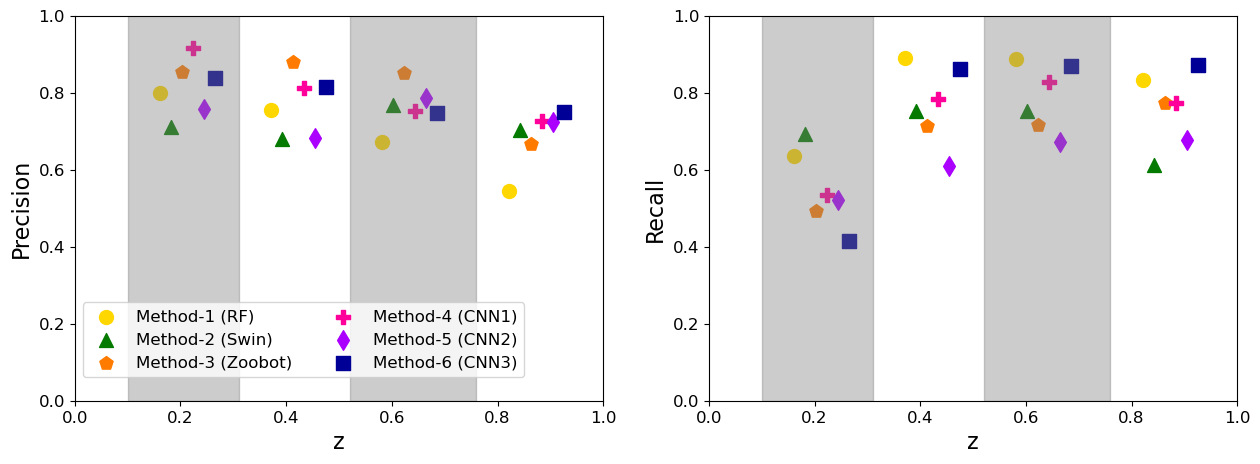}
    \caption{Precision (left) and recall (right) as a function of redshift, using HSC visual classifications of major mergers--non-mergers as true labels. For all methods, precision is higher than for the TNG training sample because in this case there is a clearer distinction between the two classes. All methods were trained on TNG and then applied to the HSC dataset.}\label{fig_prec_recall_HSC}
\end{figure*}

\begin{figure}
    \centering
    \includegraphics[width=0.48\textwidth]{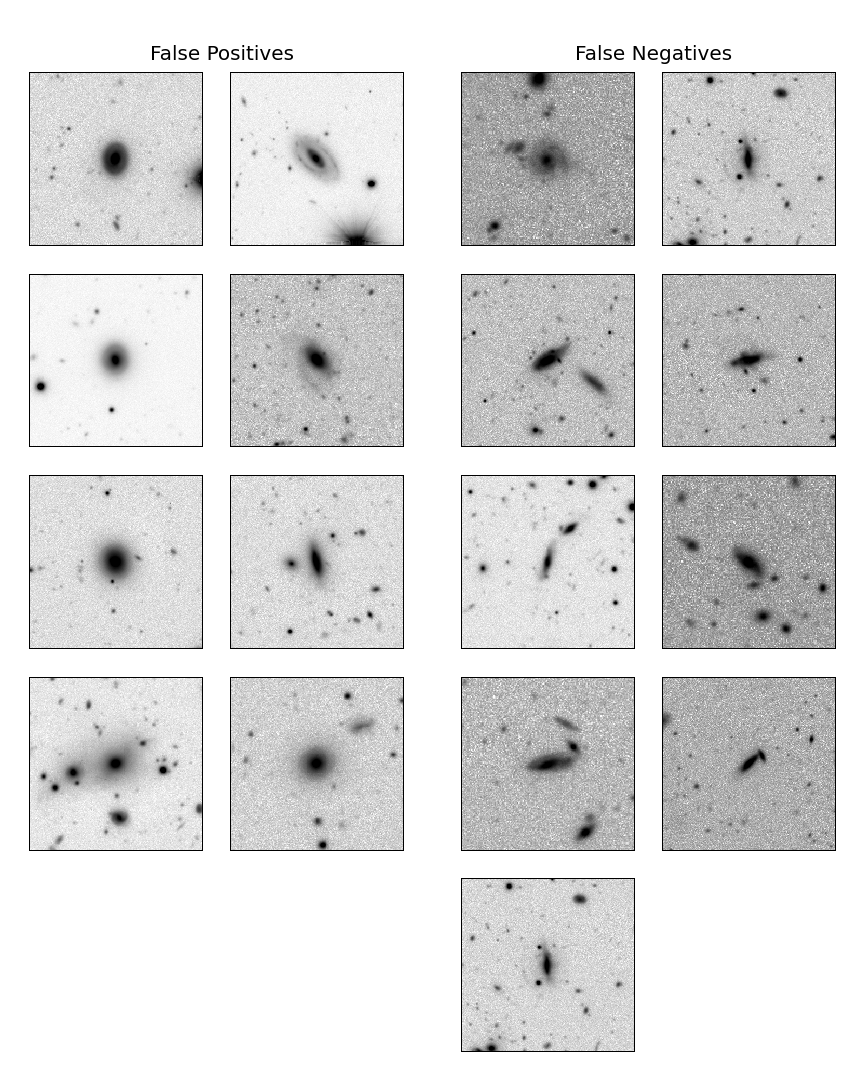}
    \caption{Galaxies for which all methods (trained on TNG) predict the wrong class when applied to HSC. False positives (left) are those visually classified as non-mergers, but that all methods predict as mergers. False negatives (right) are galaxies visually classified as mergers, while all methods predict them as non-mergers.}\label{fig.examples.fail}
\end{figure}

\begin{figure}
    \centering
    \includegraphics[width=0.48\textwidth]{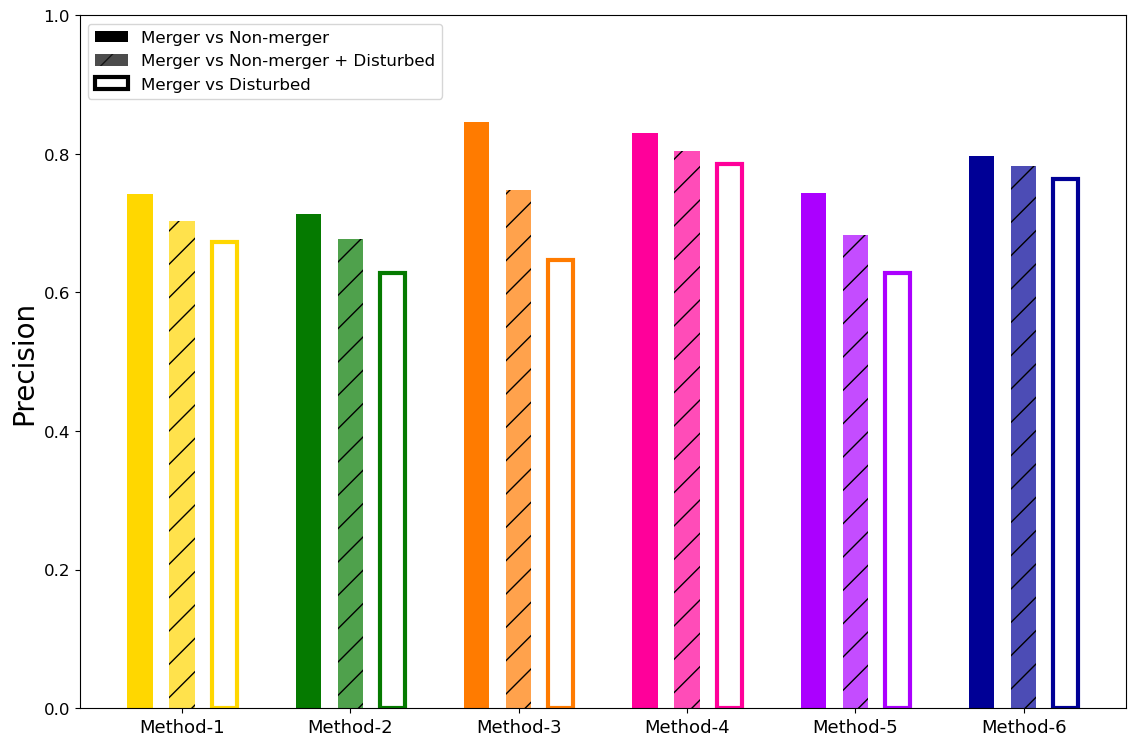}
    \caption{Precision for the HSC set using visual labels. Different definitions of negative class (non-merger): visually classified non-mergers (clear separation between the classes), visually classified non-merger plus disturbed galaxies, and disturbed galaxies. The precision for all methods drops when the separation between the classes is smaller. The recall is not shown as by definition it does not vary. All methods were trained on TNG and then applied to the HSC dataset.}\label{fig_prec_HSC}
\end{figure}

\begin{figure*}
    \centering
    \includegraphics[width=0.9\textwidth]{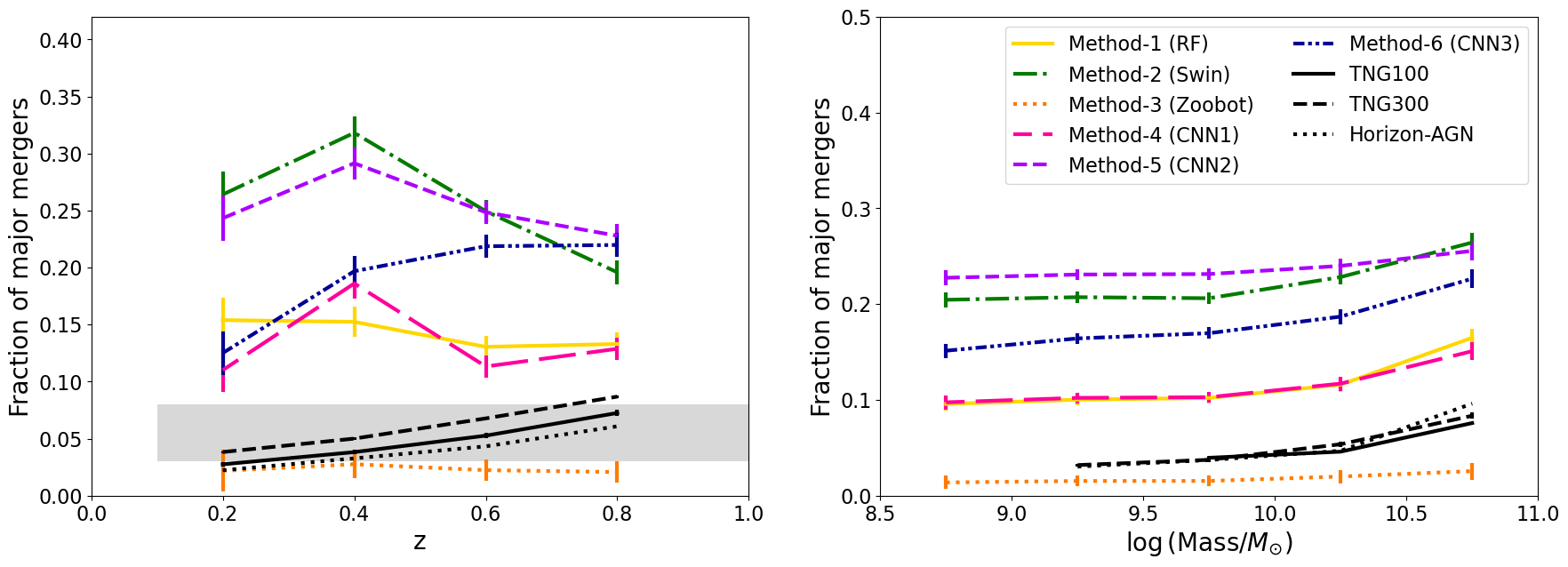}
    \caption{Fraction of major mergers as a function of redshift (left) and stellar mass (right), for each method on real HSC observations. The solid (dashed) black line shows the fraction of major mergers found in TNG100 (TNG300). The dotted black line shows the fraction of mergers for Horizon-AGN. All methods were trained on TNG and then applied to the HSC dataset.}\label{frac_merger_HSC}
\end{figure*}

\begin{table}
\caption{Performance metrics as percentages (accuracy, precision, recall, and F1-score for the merger class and AUC) of the different methods (trained on TNG) on the visual HSC set, for the binary classification task. The best performance in each metric is highlighted in bold.}\label{tab.hsc_all_redshifts}
    \centering
    \begin{tabular}{l|ccccc}
    \hline
              & \multicolumn{4}{c}{All redshift bins combined}     \\
    \hline
    Methods   & Acc.  & P     & R     & F1    & AUC \\
    \hline
    Method-1 (RF) & 74.4  & 74.1  & \bf{74.9}  & \bf{74.5}  & 80.0 \\
    Method-2 (Swin) & 71.3  & 71.4  & 71.2  & 71.3  & 79.7 \\
    Method-3 (Zoobot) & 74.1  & \bf{84.6}  & 58.8  & 69.4  & 83.4 \\
    Method-4 (CNN1) & \bf{75.5}  & 83.0  & 64.2  & 72.4  & \bf{84.7} \\
    Method-5 (CNN2) & 69.1  & 74.4  & 58.4  & 65.4  & 74.0 \\
    Method-6 (CNN3) & 72.6  & 79.9  & 60.5  & 68.9  & 80.3 \\
    \hline
    \end{tabular}
\end{table}

\begin{table*}
\caption{Summary of the performance metrics as percentages (accuracy, precision, recall, and F1-score for the merger class) of the different methods (trained on TNG) on the visual HSC set,  considering different scenarios for the non-merger class. The first scenario includes both clear non-mergers and disturbed galaxies. The second scenario only includes disturbed galaxies. The best performance in each metric is highlighted in bold.}\label{tab.hsc_disturbed}
    \centering
    \begin{tabular}{l|ccccc|ccccc}
    \hline
              & \multicolumn{4}{c}{Non-merger+disturbed}  & \multicolumn{4}{c}{Only disturbed}    \\
    \hline
    Methods   & Acc.  & P     & R     & F1  & AUC  & Acc.  & P     & R     & F1  & AUC\\
    \hline
    Method-1 (RF) & 71.4  & 70.3  & \bf{74.1}  & \bf{72.2} & 78.7 & 69.1  & 67.2  & \bf{74.5}  & 70.7 &  75.2\\
    Method-2 (Swin) & 68.5  & 67.7  & 70.6  & 69.1 & 74.3 & 64.1  & 62.9  & 69.0  & 65.8 & 65.8 \\
    Method-3 (Zoobot) & 69.6  & 75.1  & 58.7  & 65.9 & 78.3 & 63.7  & 64.8  & 59.9  & 62.2 & 69.4 \\
    Method-4 (CNN1) & \bf{74.2}  & \bf{80.4}  & 64.0  & 71.3 & \bf{81.6} & \bf{73.3}  & \bf{78.5}  & 64.1  & \bf{70.6} & \bf{78.3} \\
    Method-5 (CNN2) & 65.4  &  68.2 & 57.5  & 62.4 & 70.0 & 61.6  & 62.8  & 57.3  & 59.9 &  66.3 \\
    Method-6 (CNN3) & 71.8  & 78.3  & 60.3  & 68.1 & 78.7 & 71.4  & 76.4  & 61.9  & 68.4 &  77.2 \\
    \hline
    \end{tabular}
\end{table*}

\subsection{HSC}\label{subsec.results.hsc}

In this section, we explore the application of the models in real observations from HSC. To evaluate the performance on the HSC dataset we cannot rely on true labels, as they do not exist. Instead, we compare the classification results from different methods using visual labels. Obviously, the performance obtained in this way cannot be directly compared with the ones in previous sections, as visual inspection is biased towards the most conspicuous mergers and non-mergers. Nonetheless, we expect a good classifier trained on simulations to be able to correctly classify the majority of the most obvious mergers and non-mergers.

We use the visual classification labels from the subsample of $\sim$2000 galaxies from \citet{Goulding2018} that fall into one of the two categories corresponding to major mergers and disturbed galaxies (possible minor mergers in some cases), along with the sample of $\sim$1000 visually identified clear non-mergers in this work. Examples of these three classes can be found in Fig. \ref{fig.example_visual}. We first investigate how the methods classify the galaxies with visual labels of major mergers and non-mergers, and use these labels as if they were true labels.  Figure \ref{plot_roc_HSC} shows the ROC for all six methods, with Method-3 (Zoobot) having the overall best performance, in terms of ROC and AUC. The computed performance metrics for each method, after combining all redshift bins, are shown in Table \ref{tab.hsc_all_redshifts} and per redshift bin in Table \ref{tab.hsc_redshift_bins} in Appendix \ref{appendix.results.redshift}.  Overall, precisions are high and even higher than for TNG and Horizon AGN for all methods which is expected as this classification task should be easier (due to the greater distinction between the two visually identified classes). Precision for the whole sample ranges from $\sim71\%$ and $\sim85\%$, depending on the method. Recall, however, is generally lower than for TNG (but higher than for Horizon-AGN), ranging from $\sim58\%$ and $\sim75\%$. This means that a significant fraction of these obvious mergers are misclassified as non-mergers. In Fig. \ref{fig_prec_recall_HSC} we plot precision and recall as a function of $z$. As in Fig. \ref{fig_prec_recall_TNG} we see a downward trend with increasing $z$ for precision. Recall, on the other hand, increases with  $z$ for most methods, with the exception of Method-3 (Zoobot) which shows a downward trend. 
In Fig. \ref{fig.examples.fail} we show the galaxies for which all methods fail: False positives are galaxies visually classified as non-mergers for which all methods predict them as mergers and false negatives are galaxies visually classified as mergers but with a non-merger label according to all methods. There are only 17 galaxies for which all methods fail. In the case of the false negatives, they are not among the most obvious mergers, in general with faint features, while in the case of false positives, there tends to be a small or not obviously interacting companion. 

Visual classifications of mergers--non-mergers do not represent the real universe, as they are likely to only include the most obvious mergers and non-mergers. In the real universe, not all galaxies will fall into either category. There will be galaxies that are neither clear mergers nor clear non-mergers (visually). Therefore the results show the best-case scenario or upper limits for the performance. In order to explore this further, we use galaxies that have been visually classified as disturbed (which could include minor mergers). Galaxies in this class are expected to not be mergers by our definition, but they are neither clear mergers nor clear non-mergers visually. If we compare the performance of the models using this disturbed class as the non-mergers, it will represent the worst-case scenario. However, this probably does not reflect the real universe, as not all galaxies fall in a disturbed, but not major-merger category. That is why we then combine the clear non-mergers with these disturbed galaxies in the non-merger class. This represents a more realistic worst-case scenario. The results of these two scenarios are summarised in Table \ref{tab.hsc_disturbed}. The predictions for the scenario in which we compare mergers versus non-merger+disturbed, the precisions range from around 62\% to 81\% (for the merger class). When considering only disturbed galaxies as the non-merger class, the precision drops a further 5\%, on average. In Fig. \ref{fig_prec_HSC} we show the precision for each method and for the three scenarios, where it is clear how the precision drops, for all methods, when there is less separation between the classes, therefore, favouring more misclassifications. However, from the most realistic worst-case scenario (mergers vs. non-mergers+disturbed), the precision only drops $\sim5\%$ on average, for all methods, with respect to the best-case scenario (mergers vs. non-mergers) We note that, by definition (see Eq. \ref{eq_recall}), the recall will not change in any of these scenarios as neither TP (mergers correctly identified as mergers) nor FN (mergers incorrectly identified as non-mergers) change in these scenarios.

Finally, in Fig. \ref{frac_merger_HSC} we explore the fraction of mergers as a function of redshift and stellar mass for the different methods. The solid and dashed black lines, show the fraction of major mergers found in TNG100 and TNG300, respectively, from which we construct the training sample. The dotted black line shows the fraction of major mergers found in Horizon-AGN, compatible with the results from TNG. The grey band represents the range of merger fractions found in different observational studies. Most observational studies find merger fractions, on average, between 0.03 and 0.08, for galaxy samples within our redshift and stellar mass range (in agreement with the fraction of major mergers found in TNG and Horizon-AGN simulations). These studies calculate the merger fractions through galaxy pairs \citep[e.g.][]{Duncan2019, Ventou2017, Mundy2017, Man2012, Williams2011}, from morphological parameters \citep{Whitney2021} or from DL methods trained on simulations \citep{Ferreira2020}. The fraction of major mergers can vary by more than an order of magnitude between methods, and it shows very different trends with $z$. None of the methods reproduces the trend from the simulations as a function of redshift. Furthermore, all methods except Zoobot find higher major merger fractions in HSC than the average values found in the literature.

Part of these differences with previous literature results may be explained by how major merger fractions are affected by the definition of merger. For example in galaxy pair studies, the major merger fraction will increase with the maximum spatial separation adopted in the definition \citep{Ravel2009}. Furthermore, in galaxy-pair studies, the sample of major mergers used to calculate merger fractions will be biased towards what we refer as pre-mergers, while methods based on morphology, such as methods based on CAS or imaging, will preferentially find post-mergers \citep{Desmons2023}. In our study, the merger definition spans a larger timescale, and the methods are trained to select both pre-mergers and post-mergers, making a direct comparison not trivial. Sample selection can also impact major merger fractions. For example, \citep{Ravel2009}, with a luminosity-selected sample, found major merger fractions from about 0.03 and up to 0.7 depending on the galaxy pair definition.  On the other hand, all methods find a similar trend with stellar mass as TNG, and in agreement with observational studies that find an increase of major merger fraction with stellar mass \citep{Ventou2017, Ravel2009}

Based on Fig. \ref{frac_merger_HSC}, it may seem that major mergers do not have a significant role, However, as seen in Fig. \ref{fig_prec_recall_TNG}, we may expect the recall to drop significantly, particularly at higher redshift, indicating that the fraction of mergers could be much larger compared to what we observe. All methods show a very similar trend with stellar mass (an upward trend, also observed in TNG100 and TNG300), albeit with different absolute fractions. However, this trend may not reflect an intrinsic behaviour. As seen in Fig. \ref{fig_prec_recall_TNG_mass}, the massive end, tends to be more complete but less precise, while the lower-mass is more precise but less complete. This may be, in part, due to a bias on the training sample, in which the majority of the most massive galaxies ($M_{*}>10^{11}M_{\odot}$) are mergers, which may translate in the methods classifying the most massive galaxies as mergers regardless of any (or lack) of merger signatures.

\section{Discussions and conclusions}\label{sec.discussion}
In this paper, we benchmarked the performance of six ML-based merger detection methods that were trained on the same data. The training dataset (TNG-train) was constructed from IllustrisTNG by creating mock images that mimic the HSC survey in terms of PSF, resolution, filter, and sky background, but that do not include the effects of dust. We first evaluated the performance of all methods on the training data (TNG-test) and then on mock images from a different simulation (Horizon-AGN), constructed in a similar way. Finally, we used all methods to make predictions on real galaxies from HSC and compared them with visual classification labels. Our conclusions are summarised below:

\begin{itemize}
    \item[$\bullet$] When we have representative data, all methods achieve fairly good performance in the binary merger versus non-merger classification task, with precision $\sim$70-80\% and recall $\sim$70-77\%. For the binary task, the best overall method (Method-3, Zoobot) in terms of AUC has a precision of 80\% and recall of 74\% on TNG-test. Zoobot is the only method pre-trained with galaxy images, which may indicate that transfer learning seems to be important, even when the classification tasks are different. Traditional ML methods (trained on morphological and structural parameters) can be competitive in some cases, and they have the advantage of being easier and quicker to use and interpret. Method-4, -5, and -6 use similar CNNs, but differ in the specific architecture, image pre-processing, and augmentation. Their performance varies by a few percentage points, indicating that a given method can be improved further by fine-tuning the various aspects. Interestingly, the performance in the binary classification does not decrease much with increasing redshift ($\sim$5\% decrease in precision and $\sim$9\% decrease in recall). Stellar mass has a bigger impact, with lower precision towards higher-mass galaxies and rapidly decreasing recall towards lower-mass galaxies.    
    \item[$\bullet$] The multi-class classification task is, as expected, much more challenging for all methods, as we try to discern more subtle differences. It is particularly difficult to distinguish between ongoing-mergers and post-mergers (which can have similar features). In addition, due to the relatively short timescale of ongoing- and post-mergers, the number of galaxies in these two categories is very small, making it harder for the classifiers to learn their representation. Most methods find it easier to classify pre-mergers. Method-3 (Zoobot) has very high precision for post-mergers. 
    \item[$\bullet$] When we apply the trained classifiers to the second simulation (Horizon-AGN), we obtain similar precision to TNG with a slightly bigger difference at higher redshift. However, recall in Horizon-AGN is much worse compared to TNG, particularly at high redshift and low stellar masses. This may be due to the intrinsic differences between the two simulations, such as the effective resolution or the sub-grid physics, resulting in different types of mock images of mergers. The implication is that we can classify with high precision the type of mergers that are present in the two simulations, but different types of mergers that are not included in the training cannot be identified as easily. It is important to realise that this may be the case when we apply simulation-trained classifiers to real observations leading to detected merger samples which can be very incomplete. 
    \item[$\bullet$] When comparing the model's predictions to visual classifications of clear HSC mergers and clear non-mergers (two very distinct classes), the precision ranges from $\sim$71 to 85\%, depending on the method. These values are slightly higher than those obtained for TNG-test, which is expected, as in this case the mergers are the most obvious, and there is less confusion between the classes. However, when the classes are less distinct (i.e. the non-merger class included disturbed or minor merger galaxies), the precision drops on average by 5\%.
    \item[$\bullet$] The fraction of detected major mergers in the HSC survey does not agree with the fraction of major mergers found in TNG and Horizon-AGN simulations;  most methods find higher major merger fractions in HSC than in the simulation. Moreover, the fraction of detected major mergers can differ by more than an order of magnitude among the various methods, which also exhibit very different trends with redshift. All methods show a fairly flat relation between major merger fraction and stellar mass, with a slight increase towards more massive galaxies. The increase in the fraction of major mergers with stellar mass is also observed in TNG and Horizon-AGN simulations. However, it is not straightforward to translate this observed trend to an intrinsic relation, because of the competing effect between precision and recall. Our work demonstrates that without a detailed quantitative understanding of precision and recall it is very challenging to understand the role of mergers in galaxy evolution.
 \end{itemize}

Detecting mergers is a challenging task;  current methods achieve accuracies of $\sim 80\%$ at best in simulated data. Galaxy properties such as stellar mass and redshift have an impact on the performance and may introduce biases in the detected mergers. A good understanding of these dependences is extremely important, not only in terms of constructing sufficiently reliable and complete merger samples, but also in terms of recovering the intrinsic relations in merger fraction versus mass and redshift. Other properties not investigated here, such as the mass ratio between merging galaxies and gas content, could also have an impact and will be investigated in the future. In this paper we only used galaxy images in a single filter. However, it is reasonable to expect that combining images in different filters can improve performance. Better quality data in terms of depth and spatial resolution (e.g. from JWST and Euclid) should also lead to better results (particularly for distinguishing between different merger stages).

In this study we show that the performance of the classifiers trained on one simulation worsens when applied to a different simulation, and it is expected to decrease in real observations as well. Domain adaptation techniques that focus on domain-invariant learning may be a promising approach to alleviate this problem \citep{Ciprijanovi2020b}.  Without domain adaptation, we show in this work that the detected merger sample can be very incomplete. However, by training on mock images that are made to resemble as much as possible the real observations (in terms of resolution, noise, and background), we can obtain similar precision in the merger class, highlighting the importance of using realistic and representative training sample \citep{Bottrell2019, Ciprijanovic2020}.

\section*{Acknowledgements}
We would like to thank the anonymous referee for their
insightful comments that have improved the quality of this paper. This publication is part of the project `Clash of the Titans: deciphering the enigmatic role of cosmic collisions' (with project number VI.Vidi.193.113 of the research programme Vidi, which is (partly) financed by the Dutch Research Council (NWO). This work has made use of the Horizon cluster, on which the Horizon-AGN simulation was post-processed, hosted by the Institut d’Astrophysique de Paris. We warmly thank S. Rouberol for running it smoothly. We thank the Center for Information Technology of the University of Groningen for their support and for providing access to the Hábrók high-performance computing cluster. CB acknowledges support from the Forrest Research Foundation. H.D.S. acknowledges support by the PID2020-115098RJ-I00 grant from MCIN/AEI/10.13039/501100011033  and from the Spanish Ministry of Science and Innovation and the European Union - NextGenerationEU through the Recovery and Resilience Facility project ICTS-MRR-2021-03-CEFCA. W.J.P. has been supported by the Polish National Science Center project UMO-2020/37/B/ST9/00466. MW acknowledges funding from the Science and Technology Facilities Council (STFC) Grant Code ST/R505006/1. CBP acknowledges support by grant CM21\_CAB\_M2\_01 from the Program ``Garant\'{\i}a Juven\'{\i}l'' from the ``Comunidad de Madrid'' 2021.

%%%%%%%%%%%%%%%%%%%%%%%%%%%%%%%%%%%%%%%%%%%%%%%%%%
\section*{Data Availability}

\bibliographystyle{aa.bst}
\bibliography{references}

%%%%%%%%%%%%%%%%%%%%%%%%%%%%%%%%%%%%%%%%%%%%%%%%%%

\appendix

\section{Comparison between TNG and Horizon-AGN galaxy populations}\label{sec.appensix.comparison.data}

We explored some of the differences between the galaxy populations produced by TNG and Horizon-AGN simulations. These differences may arise from the sub-grid physics that each simulation implements. All morphological statistics shown here were been obtained from Statsmorph, as explained in Sec. \ref{subsec.methods.1}. In Fig. \ref{fig.diff.hist}, we show the distributions of different galaxy properties (stellar mass, circularised half-light radius and S\'ersic index) for mergers and non-mergers in both simulations. It appears that Horizon-AGN produces fewer massive galaxies than TNG, and the difference is larger for the mergers. In addition, Horizon-AGN produces, in general, smaller galaxies with lower S\'ersic indices. In Fig. \ref{fig.diff.gini.m20} we show how Horizon-AGN and TNG galaxies populate the Gini-M20 plane, with some dissimilarities between them. Therefore, Horizon-AGN and TNG produce slightly different galaxy populations, which may play a role in the differences observed in the performance between Horizon-AGN and TNG (as discussed in Sec. \ref{subsec.results.horizon}).

\begin{figure*}
    \centering
    \includegraphics[width=0.95\textwidth]{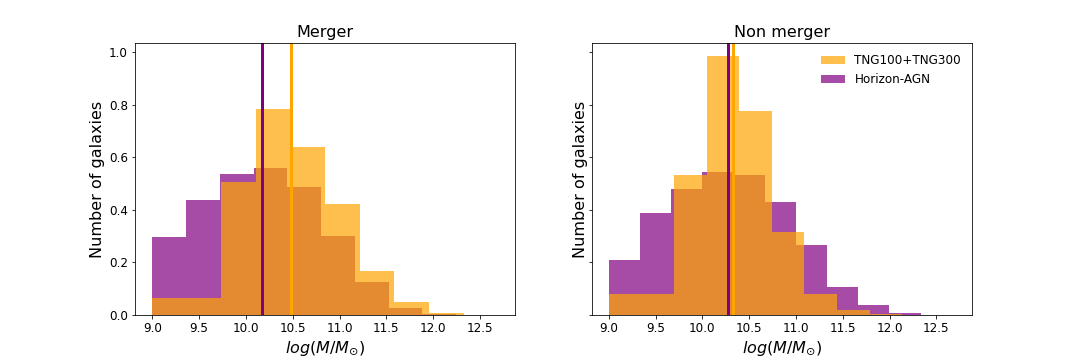}
    \includegraphics[width=0.95\textwidth]{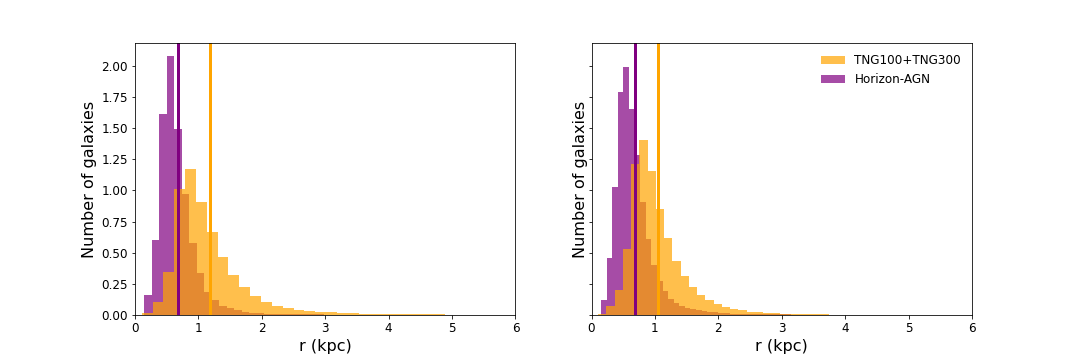}    \includegraphics[width=0.95\textwidth]{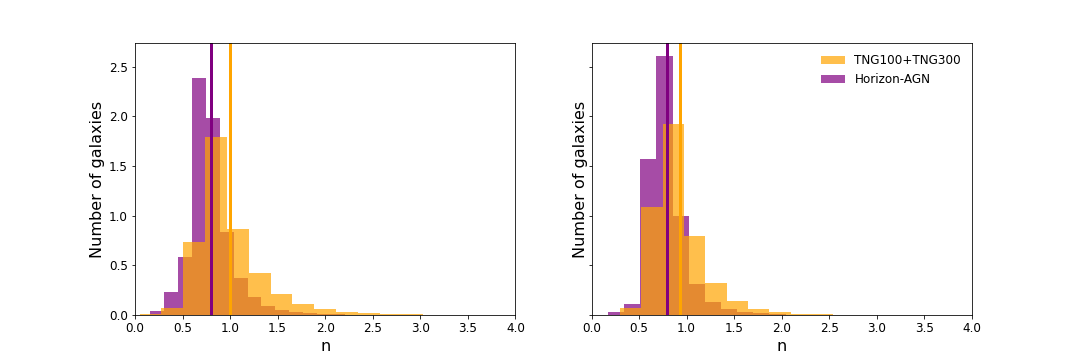}
    \caption{Stellar mass distribution, circularised half-light radius, and S\'ersic index for each simulation (combined TNG100 and TNG300, and Horizon-AGN). The left columns show the mergers (as determined by the merger trees in the simulations), and the right column shows the non-mergers. The vertical lines show the mean value of each distribution.}
    \label{fig.diff.hist}
\end{figure*}

\begin{figure*}
    \centering
    \includegraphics[width=0.95\textwidth]{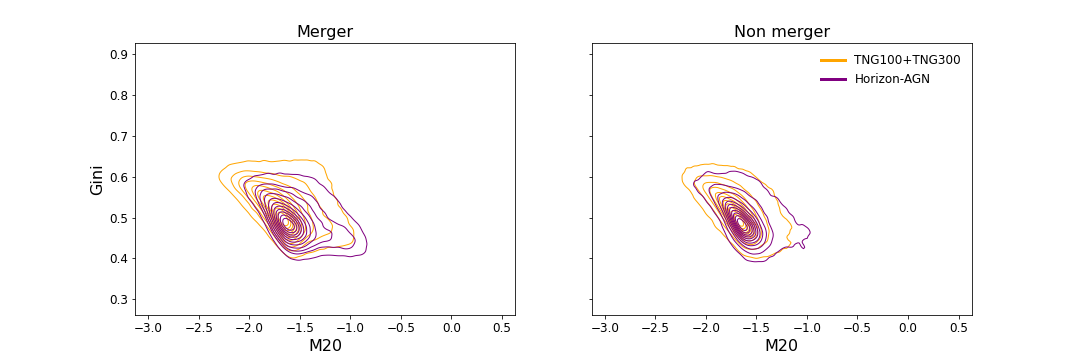}
    \caption{Gini-M20 relationship for mergers (left) and non-mergers (right), for the two simulations (TNG100 and TNG300 combined in yellow, and Horizon-AGN in purple).}
    \label{fig.diff.gini.m20}
\end{figure*}

\section{Summary of the methods}\label{appendix_summary_table}

Table \ref{tab.summary.methods} summarises the methods described in Sect. \ref{sec.methods}, to show the main differences between them. In particular, we present the name of each architecture, the number of trainable parameters, whether one single network was trained on the whole dataset or four networks were used (one for data in each redshift bin), the final size of the images (in physical size and in pixels), the scaling applied, data augmentation and any extra data that was used. Finally, we give a reference for each method.

\addtolength{\tabcolsep}{-0.3em}
\begin{sidewaystable*}
\caption{Summary of the ML and DL methods: Name or architecture, number of trainable parameters, whether one network is trained in the whole dataset or four networks are trained (one for each redshift bin), the size of the image in pixels, the size of the image in kpc, the scaling used in the images, the threshold used for the binary classification task, the kind of data augmentation used during training, any extra data that was used, whether the method performed the multi-class classification task, and a paper reference for the method.}\label{tab.summary.methods}
    \centering
    % \small
    \begin{tabular}{l|c|c|c|c|c|c}
    \hline
     & Method-1 (RF)  & Method-2 (Swin)   & Method-3 (Zoobot) & Method-4 (CNN1) & Method-5 (CNN2) & Method-6 (CNN3) \\
    \hline
    \hline
    Architecture      & RF & SwinTransformer & Pretrained-CNN  & CNN & CNN   & CNN  \\
          &  &  & (EfficientnetB0)  & (4conv+2dense) &  (4conv+2dense)  &  (3conv+1dense) \\
    \hline
    Trainable        & $2\times10^8$  (binary task)   & $2\times10^5$  &  $5.3\times10^6$  & $0.1<z<0.31$: $4.6\times10^6$ & $8.1\times10^6$ & $6.1\times10^5$ \\
    parameters        & $8\times10^7$ (multiclass task)  &   &   & $0.31<z<0.52$: $3.7\times10^6$  &  &  \\
            &   &   &   & $0.52<z<0.76$: $3.0\times10^6$  &  &  \\
            &   &   &   & $0.76<z<1.0$: $2.5\times10^6$  &  &  \\
    \hline
    4 bins together?  & Yes & Yes & Yes & No & Yes & Yes \\
    \hline
    Size (pixels)     & 310/192/160/128 & 224 & 200 & 310/192/160/128 & 128 & 100 \\
    \hline
    Size (kpc)        & 160 & 56/93/112/140 & 100 & 80  & 160 & 160 \\
    \hline
    Scaling           & None & linear& arcsinh & arcsinh+clip & linear & AsinStrech \\
               &  &  [0,1] &  & [0,1] &   [0,1] & +Percentile(97) \\
    \hline
    Threshold        & 0.5 & - & 0.56\footnote{\label{note1} The threshold was chosen to maximise TPR and minimise FPR } & 0.5 & 0.51\footnote{The threshold was chosen to maximise TPR and TNR, which is the threshold with TPR=TNR}    &  0.51\textsuperscript{\ref{note1}}\\
    \hline
    Data augm.        & --   &  Rotation        &  Rotation,   & Rotation ($0^\circ$--\ $90^\circ$), & -- & Rotation ($-45^\circ$--\ $45^\circ$)  \\
                      &    & ($0^\circ$, $90^\circ$, $180^\circ$, or $270^\circ$),      & horizontal flip, &  horizontal flip,  & & horizontal flip,  \\
                      &    & horizontal flip,       &  vertical flip    &  vertical flip & & vertical flip,  \\
                      &    & vertical flip          &    &  zoom [0.7,1.1] & & zoom [0.75,1.3], \\
                      &    &                        &                   & & &  xy translation [-0.05, 0.05)  \\
    \hline
    Extra data        & No & Yes (Pretrained    & Yes  & No & No &  No  \\
                      &    & ImageNet-1k)       & (Galaxy Zoo)     &    & &    \\
    \hline
    multi-class       & Yes & Yes & Yes & Yes & No & No \\

    \hline
    reference & \cite{Guzman-Ortega2023} & \cite{Minghao2021} & \cite{Walmsley2023} & \cite{Bickley2021} & Chudy et al. in prep. & \cite{Walmsley2019} \\
        \hline
    \end{tabular}
\end{sidewaystable*}

\section{Performance of ML/DL-methods as a function of redshift}\label{appendix.results.redshift}

Here we summarise the performance of the methods for the different datasets as a function of redshift. In Table \ref{tab.tng_redshift_bins} we show the performance (accuracy, precision, recall, and F1-score) of all six methods on the TNG-test dataset. As discussed in Sec \ref{subsec.results.tng} Zoobot shows the overall best performance over the whole redshift range. Table \ref{tab.tng_multi_redshift_bins} shows the precision and recall for pre-mergers and post-mergers as a function of redshift. Tables \ref{tab.horizon_redshift_bins} and \ref{tab.hsc_redshift_bins} show the performance obtained on Horizon-AGN and HSC, respectively, as a function of redshift.

\begin{table*}
\caption{Performance metrics as percentages (accuracy, precision, recall, and F1-score for the merger class) of the different methods on the TNG-test set per redshift bin, for the binary classification task. The best performance in each metric is highlighted in bold.}\label{tab.tng_redshift_bins}
    \centering
    \small
    \begin{tabular}{l|cccc|cccc|cccc|cccc}
    \hline
              &    \multicolumn{4}{c|}{$0.1<z<0.31$}     &  \multicolumn{4}{c|}{$0.31<z<0.52$}    &  \multicolumn{4}{c|}{$0.52<z<0.76$}      &   \multicolumn{4}{c}{$0.76<z<1.0$}     \\
    \hline
    Methods   & Acc.  & P     & R     & F1    & Acc.  & P     & R     & F1    & Acc.  & P     & R     & F1    & Acc.  & P     & R    & F1    \\
    \hline
    M-1 (RF)    & 73.4  & 74.5  & 71.0  & 72.7  & 72.7  & 73.9  & 70.3  & 72.1  & 70.0  & 70.8  & 67.9  & 69.3  & 67.4  & 68.2  & 65.0  & 66.6 \\
    M-2 (Swin)  & 74.7  & 70.4  & \bf{85.2}  & 76.8  & 73.3  & 68.3  & \bf{87.1}  & 76.4  & 72.4  & 68.1  & \bf{84.1}  & 75.3  & 72.0  & 70.4  & \bf{76.2}  & \bf{73.2} \\
    M-3 (Zoobot)& \bf{80.8}  & \bf{84.1}  & 75.8  & \bf{79.8}  & \bf{80.3}  & \bf{82.5}  & 77.0  & \bf{79.7}  & \bf{76.8}  & \bf{78.6}  & 73.7  & \bf{76.1}  & \bf{73.5}  & \bf{75.7}  & 69.3  & 72.4 \\
    M-4 (CNN1)  & 77.3  & 82.5  & 70.8  & 75.7  & 76.1  & 77.0  & 75.6  & 76.3  & 74.0  & 74.9  & 73.0  & 73.9  & 70.8  & 71.4  & 68.8  & 70.1 \\
    M-5 (CNN2)  & 74.0  & 73.7  & 74.4  & 74.1  & 72.7  & 72.9  & 72.2  & 72.6  & 69.6  & 71.0  & 66.4  & 68.6  & 66.6  & 70.4  & 57.5  & 63.3 \\
    M-6 (CNN3)  & 74.2  & 76.0  & 70.7  & 73.3  & 73.4  & 73.9  & 72.3  & 73.1  & 70.9  & 72.3  & 67.7  & 70.0  & 68.1  & 71.3  & 60.8  & 65.6 \\
    \hline
    \end{tabular}
\end{table*}

\begin{table*}
\caption{Performance metrics as percentages (precision and recall for the pre-mergers and post-mergers) of the different methods on the TNG-test set per redshift bin, for the multi-class classification task. The post-merger class includes the ongoing-mergers. The merger classes or stages are determined by the merger trees from the TNG simulation. The best performance in each metric is highlighted in bold.}\label{tab.tng_multi_redshift_bins}
    \centering
    \small
    \begin{tabular}{l|cccc|cccc|cccc|cccc}
    \hline
              &    \multicolumn{4}{c|}{$0.1<z<0.31$}     &  \multicolumn{4}{c|}{$0.31<z<0.52$}    &  \multicolumn{4}{c|}{$0.52<z<0.76$}      &   \multicolumn{4}{c}{$0.76<z<1.0$}     \\
    \hline
              &    \multicolumn{2}{c|}{Pre-merger}  &    \multicolumn{2}{c|}{Post-merger}   &    \multicolumn{2}{c|}{Pre-merger}  &    \multicolumn{2}{c|}{Post-merger}   &    \multicolumn{2}{c|}{Pre-merger}  &    \multicolumn{2}{c|}{Post-merger}   &    \multicolumn{2}{c|}{Pre-merger}  &    \multicolumn{2}{c}{Post-merger}     \\
    \hline
    Methods   & P  & \multicolumn{1}{c|}{R}     & P     & R    & P  & \multicolumn{1}{c|}{R}     & P     & R    & P  & \multicolumn{1}{c|}{R}     & P     & R    & P  & \multicolumn{1}{c|}{R}     & P     & R     \\
    \hline
    M-1 (RF)    & 62.9  & 48.5  & 56.2  & 68.0  & 63.2  & 45.9  & 57.1  & 65.9  & 58.7  & 40.1  & 53.7  & 65.0  & 53.8  & 36.8  & 51.3  & 59.5 \\
    M-2 (Swin)  & \bf{76.6}  & 66.3  & 61.0  & 73.3  & \bf{75.8}  & 69.4  & 59.4  & 70.9  & \bf{70.5}  & 66.5  & 59.4  & 64.0  & \bf{65.6}  & 60.8  & 61.1  & 52.8 \\
    M-3 (Zoobot)& 69.5  & \bf{76.3}  & \bf{85.6}  & 42.1  & 68.6  & \bf{76.8}  & \bf{83.0}  & 43.8  & 65.0  & \bf{75.3}  & \bf{77.5}  & 35.8  & 59.0  & \bf{68.9}  & \bf{77.6}  & 30.7 \\
    M-4 (CNN1)  & 68.8  & 29.8  & 41.1  & \bf{85.2}  & 67.6  & 23.3  & 41.3  & \bf{83.4}  & 59.1  &  8.9  & 41.5  & \bf{73.5}  & 60.5  & 12.4  & 43.6  & \bf{66.5} \\
    \hline
    \end{tabular}
\end{table*}

\begin{table*}
\caption{Performance metrics as percentages (accuracy, precision, recall, and F1-score for the merger class) for the different methods (trained on TNG) applied to the Horizon-AGN set per redshift bin, for the binary classification task. The best performance in each metric is highlighted in bold.}\label{tab.horizon_redshift_bins}
    \centering
    \small
    \begin{tabular}{l|cccc|cccc|cccc|cccc}
    \hline
              &    \multicolumn{4}{c|}{$0.1<z<0.31$}     &  \multicolumn{4}{c|}{$0.31<z<0.52$}    &  \multicolumn{4}{c|}{$0.52<z<0.76$}      &   \multicolumn{4}{c}{$0.76<z<1.0$}     \\
    \hline
    Methods      & Acc.  & P     & R     & F1    & Acc.  & P     & R     & F1    & Acc.  & P     & R     & F1    & Acc.  & P     & R    & F1    \\
    \hline
    M-1 (RF)    & 60.3  & 78.5  & 28.4  & 41.7  & 58.5  & \bf{72.8}  & 27.1  & 39.5  & 57.4  & \bf{67.3}  & 28.8  & 40.3  & 52.1  & \bf{57.7}  & 16.0  & 25.1 \\
    M-2 (Swin)  & \bf{71.0}  & 69.1  & \bf{76.2}  & \bf{72.4}  & \bf{65.9}  & 70.9  & \bf{53.9}  & \bf{61.2}  & \bf{59.7}  & 65.9  & 40.1  & 49.9  & 52.9  & 61.2  & 16.1  & 25.5 \\
    M-3 (Zoobot)& 66.2  & 83.4  & 40.4  & 54.4  & 59.1  & 72.6  & 29.4  & 41.8  & 54.2  & 62.9  & 20.4  & 30.8  & 51.3  & 57.4  &  10.2  & 17.3 \\
    M-4 (CNN1)  & 65.4  & \bf{84.3}  & 37.9  & 52.2  & 59.8  & 72.9  & 31.2  & 43.7  & 53.6  & 60.9  & 20.0  & 30.2  & 50.9  & 56.4  &  7.6  & 13.5 \\
    M-5 (CNN2)  & 70.6  & 72.4  & 66.6  & 69.4  & 63.1  & 66.6  & 52.8  & 58.9  & 58.1  & 61.7  & \bf{42.6}  & \bf{50.4}  & \bf{53.5}  & 56.8  & \bf{28.9}  & \bf{38.3}  \\
    M-6 (CNN3)  & 55.8  & 73.5  & 18.2  & 29.2  & 54.6  & \bf{72.8}  & 14.7  & 24.4  & 52.3  & 62.2  & 11.7  & 19.8  & 50.9  & 57.6  &  6.5  & 11.7 \\
    \hline
    \end{tabular}
\end{table*}

\begin{table*}
\caption{Performance metrics as percentages (accuracy, precision, recall, and F1-score for the merger class) of the different methods (trained on TNG) on the HSC set per redshift bin, for the binary classification task. The best performance in each metric is highlighted in bold.}\label{tab.hsc_redshift_bins}
    \centering
    \small
    \begin{tabular}{l|cccc|cccc|cccc|cccc}
    \hline
              &    \multicolumn{4}{c|}{$0.1<z<0.31$}     &  \multicolumn{4}{c|}{$0.31<z<0.52$}    &  \multicolumn{4}{c|}{$0.52<z<0.76$}      &   \multicolumn{4}{c}{$0.76<z<1.0$}     \\
    \hline
    Methods      & Acc.  & P     & R     & F1    & Acc.  & P     & R     & F1    & Acc.  & P     & R     & F1    & Acc.  & P     & R    & F1    \\
    \hline
    M-1 (RF)     & 73.8  & 80.0  & 63.5  & \bf{70.8}  & 80.1  & 75.6  & \bf{88.9}  & 81.7  & 72.7  & 67.2  & \bf{88.9}  & 76.5  & 56.7  & 54.3  & 83.3  & 65.8 \\
    M-2 (Swin)   & 70.7  & 71.2  & \bf{69.4}  & 70.3  & 69.9  & 68.0  & 75.2  & 71.4  & 76.3  & 76.8  & 75.3  & 76.0  & 67.7  & 70.4  & 61.3  & 65.5 \\
    M-3  (Zoobot)& 70.4  & 85.4  & 49.3  & 62.5  & 80.8  & \bf{88.0}  & 71.2  & 78.7  & \bf{79.5}  & \bf{85.0}  & 71.7  & 77.8  & 69.4  & 66.7  & 77.4  & 71.6 \\
    M-4 (CNN1)   & \bf{74.3}  & \bf{91.7}  & 53.5  & 67.6  & 80.1  & 81.2  & 78.3  & 79.7  & 77.8  & 75.2  & 82.8  & 78.8  & 74.2  & 72.7  & 77.4  & 75.0 \\
    M-5 (CNN2)   & 67.7  & 75.7  & 52.0  & 61.7  & 66.4  & 68.3  & 61.1  & 64.5  & 74.5  & 78.7  & 67.2  & 72.5  & 71.0  & 72.4  & 67.7  & 70.0 \\
    M-6 (CNN3)   & 66.8  & 83.9  & 41.5  & 55.5  & \bf{83.4}  & 82.6  & 86.3  & \bf{83.9}  & 78.8  & 74.8  & 86.9  & \bf{80.4}  & \bf{79.0} & \bf{75.0}  & \bf{87.1}  & \bf{80.6} \\
    \hline
    \end{tabular}
\end{table*}

\section{Merger stages: Four classes}\label{appendix.4classes}

Here we present the results for the 4-class classification for methods 1-4. The methods are trained to classify galaxies into four classes: non-merger, pre-merger, ongoing-merger, and post-merger. These four merger stages have been determined from the merger trees from the TNG simulation. Figure \ref{fig.confusion.4class.tng} shows the confusion matrices for the four methods for all redshift bins combined. The matrices are normalised to show the precision of each class (with recall values shown in brackets). The performance is relatively low, with only some methods achieving more than 50\% precision for some of the classes. In particular, it is clear that all methods are failing at distinguishing between ongoing-mergers and post-mergers. This may be because these two classes have the lowest number of galaxies, and they can present very similar features. In Sect. \ref{subsubsec.results.tng.multi}, we show the performance of the methods when combining these two classes in one.

\begin{figure*}
    \centering
    \includegraphics[width=0.4\textwidth]{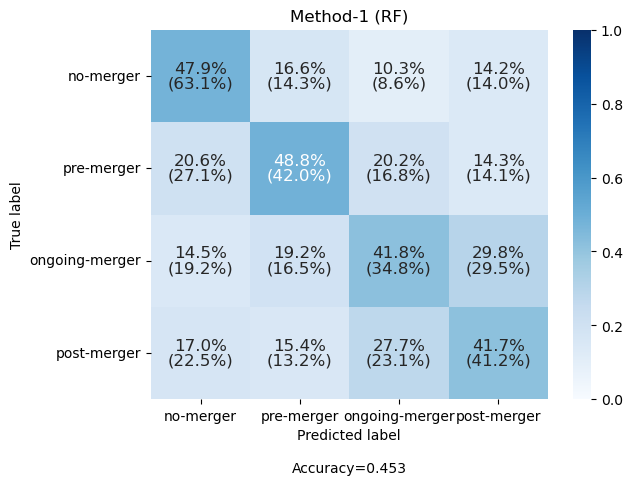}
    \includegraphics[width=0.4\textwidth]{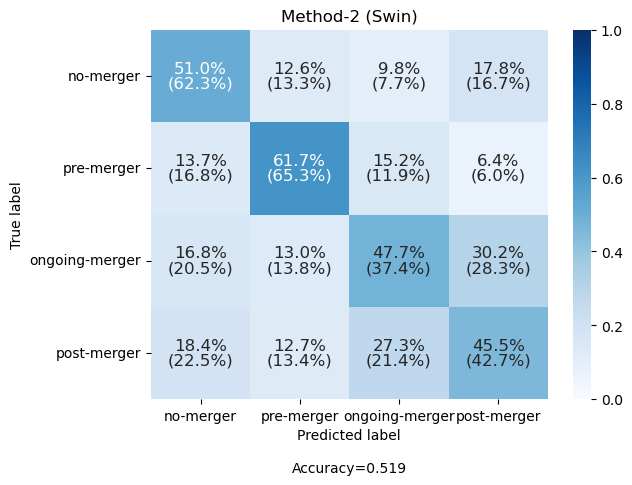}
    \includegraphics[width=0.4\textwidth]{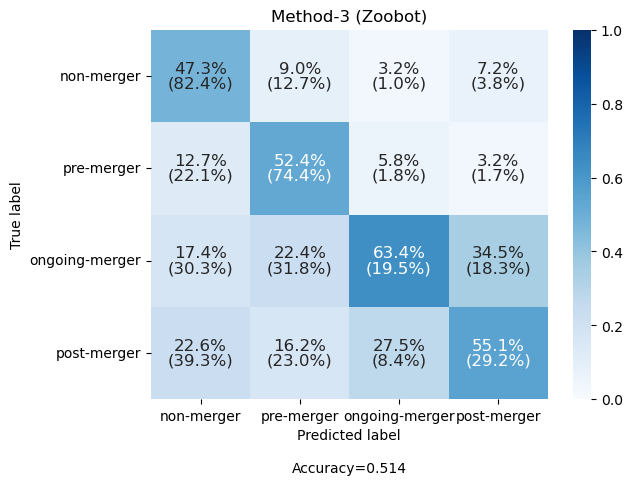}
    \includegraphics[width=0.4\textwidth]{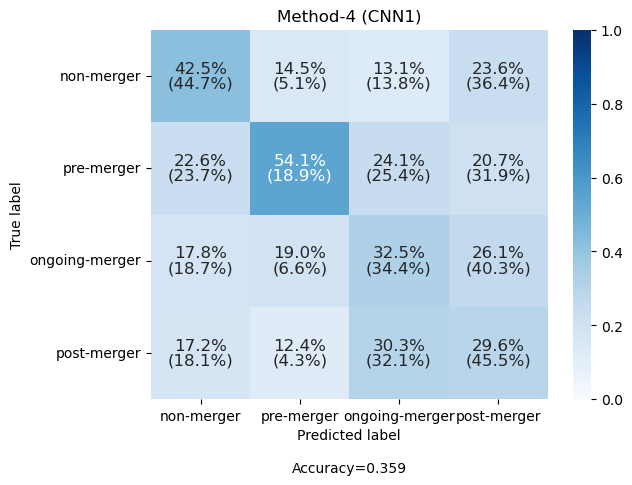}
    \caption{Confusion matrices for Method-1 (top left panel), Method-2 (top right panel), Method-3 (bottom left panel), and Method-4 (bottom right panel) for the four-class classification task on TNG. The four classes are non-merger, pre-merger, ongoing-merger, and post-merger (the different classes determined by the merger trees in the simulation). The data from all four redshift bins are combined. The confusion matrices are normalised vertically; therefore, the diagonal represents the precision of each class.
The recall is shown in brackets.}\label{fig.confusion.4class.tng}
\end{figure*}

\end{document}